\documentclass[12pt]{article}
\RequirePackage{setspace}
    \AtBeginDocument{\setlength\abovedisplayskip{3pt}}
    \AtBeginDocument{\setlength\belowdisplayskip{3pt}}
\usepackage{amsthm}
\usepackage{amsfonts}
\usepackage{bbm}
\usepackage{graphicx}
\usepackage{sectsty}
\sectionfont{\normalsize}
\usepackage{appendix}
\usepackage{float}
\usepackage{url}
\usepackage{caption}
\usepackage[round]{natbib}

\begin{document}
\title{\Large The unreasonable success of quantum probability I:\\
\large Quantum measurements as uniform fluctuations}
\author{\normalsize Diederik Aerts \\ 
        \small\itshape \vspace{-0.1 cm}
        Center Leo Apostel for Interdisciplinary Studies and Department  \\
        \small\itshape \vspace{-0.1 cm}
        of Mathematics, Brussels Free University, Brussels, Belgium \\ 
         \small \vspace{-0.1 cm}
        email: \url{diraerts@vub.ac.be} \vspace{0.2 cm} \\ 
             \normalsize   Massimiliano Sassoli de Bianchi  \\ 
       \vspace{-0.1 cm} \small\itshape
        Laboratorio di Autoricerca di Base,
         Lugano, Switzerland \\
        \small \vspace{-0.1 cm}
        email: \url{autoricerca@gmail.com} \vspace{0.2 cm}
        }
\date{}
\maketitle
\vspace{-1cm}
\begin{abstract}
\noindent 
We introduce a model  which allows to represent the probabilities associated with an arbitrary measurement situation as it appears in different domains of science  -- from cognitive science to physics  -- and use it to explain the emergence of quantum probabilities (the Born rule) as \emph{uniform} fluctuations on this measurement situation. The model exploits the geometry of simplexes to represent the states both of the system and the measuring apparatus, in a way that the measurement probabilities can be derived as the Lebesgue measure of suitably defined convex subregions of the simplex under consideration. Although  the model we propose, which we call the \emph{uniform tension-reduction} (UTR) model,
is an abstract construct, it admits physical realizations. In this article we consider a very simple and evocative one, using a material point particle which is acted upon by special elastic membranes, which by breaking and collapsing are able to ``release the tension'' and produce  the different possible outcomes. This easy to visualize mechanical realization allows one to gain considerable insight into the possible hidden structure of a measurement process, be it from a measurement associated with a situation in cognitive science or in physics, or in any other domain. We also show that the UTR-model can be further generalized into a model describing conditions of lack of knowledge generated by \emph{non-uniform} fluctuations, which we call the \emph{general tension-reduction} (GTR) model. In this more general framework,  which is more suitable to describe typical experiments in cognitive science, we define and motivate a notion of \emph{universal measurement}, describing the most general possible condition of lack of knowledge in a measurement,  emphasizing that the uniform fluctuations characterizing quantum measurements can also be understood as an average over all possible forms of non-uniform fluctuations which can be actualized in a measurement context. This means that the Born rule of quantum mechanics can be understood as a first order approximation of a more general non-uniform theory, thus explaining part of the great success of quantum probability in the description of different domains of reality. And more specifically, also providing a possible explanation for the success of quantum cognition, a research field in cognitive science employing the quantum formalism as a modeling tool. This is the first part of a two-part article. In the second part \citep{AertsSassolideBianchi2014a}, the proof of the equivalence between universal measurements and  uniform measurements, and its significance for quantum theory as a first order approximation, is given and further analyzed.
\end{abstract}

\vspace{-0.1cm}

\vspace{-0.4cm}
\section{Introduction}
\label{Introduction}
\vspace{-0.4cm}

The great success of mathematics in the natural sciences has always amazed and enchanted scientists \citep{Wigner1960}, and quantum mechanics, with its use of very sophisticated mathematical notions in the description of  physical entities (such as complex Hilbert spaces with an Hermitian scalar product and self-adjoint operators) is the perfect example of a theory  which has taken full advantage of an advanced mathematical language. But quantum mechanics is not only remarkable  for the sophistication of its mathematics: it is also for its  ``unreasonable'' success in the description of a vast class of phenomena, not limited to those traditionally investigated by quantum physicists. 

The most surprising application of quantum physics, beyond the domain of microphysics, is probably in the study of human cognitive processes. Indeed, the mathematical structure of quantum theory, with its non-classical (non-Kolmogorovian) probability calculus, has been used with considerable success in the past decade to model aspects of human cognition, such that a new field of research within cognitive science, referred to as `quantum cognition', emerged \citep{AertsAerts1995, AertsBroekaertGaboraSozzo2013, AertsGabora2005a, AertsGabora2005b, AertsGaboraSozzo2013, Blutner2009, Blutner2013, BruzaetBusemeyerGabora2009, Bruzaetal2007, Bruzaetal2008a, Bruzaetal2008, Bruzaetal2009, Bruzaetal2009b, BusemeyerBruza2012, Busemeyeretal2011, Busemeyeretal2006, Franco2009, HavenKhrennikov2013, GaboraAerts2002, Khrennikov2010, KhrennikovHaven2009, PothosBusemeyer2009, VanRijsbergen2004, Wangetal2013, YukalovSornette2010}.

As a matter of fact, in quite some of these `quantum cognition models,' it is shown that quantum probabilities are more adapted and effective as compared to traditional approaches -- based on classical, Kolmogorovian probabilities -- in capturing the way humans deal with their thinking through concepts and their combinations, and the way they make their decisions. 

Of course, regarding this ability of the quantum formalism in matching the description of not only microscopic entities, for the description of which it was invented and construed, but also of mental ones, as studied by cognitive and decision scientists, we can always say, quoting \citet{BusemeyerBruza2012}, that:
\vspace{-0.25cm}
\begin{quote}
``[$\cdots$] many areas of inquiry that were historically part of physics are now considered part of mathematics, including complexity theory, geometry, and stochastic processes. Originally they were applied to physical entities and events. For geometry, this was shapes of objects in space. For stochastic processes, this was statistical mechanics of particles. Over time they became generalized and applied to other domains. Thus, what happens here with quantum mechanics mirrors the history of many, if not most, branches of mathematics.''
 \end{quote}
\vspace{-0.25cm}

In other terms, we can argue that the effectiveness of quantum mechanics in other  fields of investigation is just  part of  the effectiveness of mathematics in science in general. Without a doubt, the understanding of the general link between mathematics, physics and  the human mind, is a fundamental metaphysical question, certainly worth investigating. In the present article, however, we shall only be concerned with a more modest and specific, although not less interesting, question, which is the following:  Why the quantum approach works so well in the modeling of so many systems and their interactions,  beyond the microscopic realm, and particularly the data of a great number of experiments on concepts, notably those studying combinations of concept,
and on human decision making? 

Let us recall that since the fifties of the last century some specific problems in economics, known as the Allais paradox \citep{Allais1953} and the Ellsberg paradox \citep{Ellsberg1961}, already indicated in those years the possibility of a violation, in human decision processes, of principles based on classical logic, like  the so-called expected utility hypothesis \citep{vonNeumannMorgenstern1944} and Sure-Thing Principle \citep{Savage1954}. In the eighties and nineties, psychologists studied in a focused way different types of human thought structures related to specific situations, were fallacies and effects such as the conjunction fallacy \citep{TverskyKahneman1983} and the disjunction effect \citep{TverskyShafir1992} are amongst the most well-known. One of the possible hypotheses with respect to these examples is that they constitute instances of human thought deviating from classical logical thought. 

Since then, indeed, decision researchers have discovered the value of quantum modeling, making a profitable use of quantum decision models for the description of a large number of experimentally identified effects \citep{Busemeyeretal2006, Busemeyeretal2011, LambertMogilianskyetal2009, PothosBusemeyer2009}, such as the conjunction fallacy and the disjunction effect \citep{Aerts2009, Blutner2009, Franco2009, Khrennikov2010, YukalovSornette2010}. In this regard, let us mention that an explanation of the violation of the expected utility hypothesis and the Sure-Thing Principle has now been modeled quantum cognitively, in terms of quantum interference effects \citep{Busemeyeretal2006, Franco2007, KhrennikovHaven2009, PothosBusemeyer2009}, and that quantum structures have  also proven their pertinence  in the ambit of  information retrieval research, a booming and significant domain in computer science, building on semantic space approaches \citep{VanRijsbergen2004, WiddowsPeters2003}.

Concerning concept combinations, very significant deviations from classicality were found in experiments conducted in the eighties by \citet{Hampton1988a, Hampton1988b}. One of us and his collaborators were able to recognize in these deviations from classicality for concept combinations the unmistakable signature of the presence of quantum structures, with their typical effects of interference, contextuality, entanglement and emergence \citep{Aerts2009, Aertsetal2000, AertsGabora2005a, AertsGabora2005b, AertsGaboraSozzo2013, AertsSozzo2011, GaboraAerts2002}. Also the so called `borderline contradictions', experimentally identified deviations from classical logic when a concept is conjuncted with its negation, could recently be modeled within the quantum cognition approach \citep{Blutner2013,Sozzo2014}. A key step in the elaboration of the approach of quantum structures modeling concepts as in \cite{Aerts2009,AertsGabora2005a,AertsGabora2005b,AertsGaboraSozzo2013}, was the possibility to formalize a concept as an \emph{entity in a specific state}, and  a context as a ``surrounding'' which is able to produce a change (either deterministic or indeterministic) of such state \citep{AertsGabora2005a, GaboraAerts2002}.

Just to give an example, consider the concept \emph{Pet}. When it is not under the influence of a specific context,  we can say that it is in its ground state, which can be understood as a sort of basic prototype of the concept. But as soon as the concept  \emph{Pet} is contextualized, for instance in the ambit of the phrase \emph{Did you see the type of pet he has? This explains that he is a weird person}, its state will change, so  that its previous ground state will stop playing the role of a prototype, which will now be played by its new state, as a sort of new `contextualized prototype'. 

The difference between the concept \emph{Pet} in a ground state and in an ``excited'' state, like the one associated with the above ``weird person context,'' can  be assessed by submitting the concept to an additional context: that of the mind of a human subject, when it is asked to select a good exemplar of that concept, among a number of possible choices. The difference between these two states will then manifest, for example, in the fact that exemplars like \emph{Snake} and \emph{Spider} will be chosen much more frequently (i.e., with a higher probability) when the \emph{Pet} concept is in the ``weird person'' excited state, rather than in its ground state. 

It is worth noticing that an exemplar of a concept also represents a possible state for it, although usually of a more concrete kind. For example, the \emph{Snake} exemplar corresponds to the more specific state \emph{The Pet is a snake}. Furthermore, the decision process of a human subject, when selecting a good example of a concept in a given state, is also to be understood as a  context, changing its state into a more concrete one. More precisely, one can think of the process of placing the concept \emph{Pet} in the context of other concepts (as we have done when placing it in the phrase \emph{Did you see the type of pet he has? This explains that he is a weird person}), as a determinative process, similar to what in physics is called a \emph{preparation}. 

On the other hand, when a human subject is stimulated to give a \emph{good example of} the concept in that state, for instance choosing between  \emph{Snake}, \emph{Spider} and other exemplars, then such context should be thought of as a \emph{measurement},  similar to  the  \emph{quantum measurements of the first kind} performed in modern physics' laboratories. Indeed, the process being  interrogative,  its outcomes are generally unpredictable, and the different  exemplars among which the subject has to choose define the eigenstates of the semantic observable s/he is effectively measuring. And of course, the relative frequencies of the measured outcomes of this observable will depend on the state of preparation of the  conceptual entity. 

Now, coming back to our initial question, about the relevance of the quantum formalism in the description  of human's cognitive and decision processes, as we said it was only in more recent times, in the beginning of this century, that it was possible to explicitly show, relying on previous investigation in quantum probability, that the observed violations of classical logic in Hampton's and other experiments could not, in any way, be modeled in the ambit of a classical (Kolmogorovian) probability theory, not even when fuzzy structures were allowed. This is a result with strong implications for the nature of human thought itself, as it shows that something with a genuine non-classical structure is at work, ``in the background,'' in our cognitive processes. In other terms, the success of quantum physics in describing concepts and their combinations can, at least in part, be explained by considering what is the main difference  between a non-classical (quantum, or quantum-like) probability theory, and a traditional classical one. 

This difference lies essentially in that quantum probabilities (and more generally quantum-like probabilities) typically describe conditions of lack of knowledge regarding properties that are \emph{created} during an experiment (the level of \emph{potentiality} of the system), whereas classical probabilities only describe conditions of lack of knowledge regarding  properties which were already  actual before the experiment was executed. This means that the inadequacy of a classical probability calculus (which is implicit in all traditional approaches, also those based on fuzzy-set theory) in the modeling of human cognition is due precisely to its inadequacy in describing  processes of  creation (to be understood here as processes of
\emph{actualization of potential properties}), whereas the adequacy of a quantum probability calculus is due to the fact  that it was historically designed to do precisely this. In other terms, if the  quantum approach to cognition works so well,  it is because both the `microscopic layer' of our physical reality, populated by so-called quantum ``particles,''  and  the `cognitive layer' of our mental reality, populated by conceptual entities, are realms of genuine `potentialities,' not of the type of a `lack of knowledge of actualities.'

So, there are very convincing reasons explaining why quantum physics performs so well in its modeling of human concepts, and we can say that these reasons are now beginning to be fairly well understood. Of course, much more can be said in this regard, but we refer the reader to the above mentioned references and in particular to the analysis presented in \citet{AertsGaboraSozzo2013} (see also \citet{Kitto2008} and the references cited therein), 
as our scope in the present article, and in its continuation \citep{AertsSassolideBianchi2014a}, is to concentrate on a different issue regarding the ``unreasonable'' effectiveness of the quantum mechanical formalism. What we are here referring to is that orthodox quantum theory is not the only available theory which is able to describe the level of potentiality present in a system and the associated processes of actualization of potential properties. 

To better explain what we mean, let us refer for example to the investigation in the axiomatic and operational content of physical theories, where standard quantum theory is built axiomatically starting from a much more general and as much as possible operationally founded approach. There have been many of such axiomatic quantum approaches, and it was even John von Neumann himself who instigated the domain of research  \citep{BirkhoffvonNeumann1936}. We can recall, among other pioneers of this field of foundational investigation, \citet{Mackey1963}, \citet{Jauch1968}, \citet{Piron1976} \citet{FoulisRandall1978} and \citet{Ludwig1983}. Also one of the authors, and his collaborators, were thoroughly engaged in research on quantum axiomatic in the foregoing century \citep{Aerts1982a, Aerts1986, Aerts1999a, AertsDurt1994, Aertsetal1997a, Aertsetal1999b}. In the course of these investigations they were able to identify relevant mathematical structures which are more general than those used in classical and quantum physics and, interestingly, it was also  possible to find explicit macroscopic situations, not necessarily related to the description of entities of the microworld, which were  conveniently described only by these  more general intermediary structures,  containing the pure classical and pure quantum structures as special limit cases \citep{Aerts1982b, Aerts1991, AertsVanBogaert1992, Aertsetal1993, Aertsetal2000}.

Let us mention that the possibility of using quantum-like structures of a very general kind to properly model macroscopic situations has played an essential role in providing evidence that a quantum-like approach would also be appropriate for situations in human cognition. In this regard, we can refer to the introduction of the SCoP (State Context Property) formalism -- a generalized quantum theory -- for the modeling of concepts in \citet{AertsGabora2005a}. Let us also mention the algebraic approach of  beim Graben and collegues \citep{beimetal2006,beimetal2009,beimetal2013}, in which classical dynamical systems such as neural networks are shown to be able to exhibit quantum-like properties in the case of coarse-graining measurements, when testing properties that cannot distinguish between epistemically equivalent states. Since in neuroscience most measurements, such as electroencephalography or magnetic resonance imaging, can be considered to be coarse-graining measurements in this sense, this ``epistemic quantization'' has direct implications for brain neurophysiology, and this without needing to refer to a notion of a ``quantum brain.''

However, what wasn't expected in the beginning of these investigations, is that pure Hilbertian structures, and the associated Born rule for calculating the probabilities of the outcomes, would be so effective in the modeling of the main quantum effects identified in these domains, different from the microworld \citep{AertsSozzo2012a, AertsSozzo2012b}. Therefore, the following question arises in a natural way: Why the Born rule, and not other ``rules,'' associated with more general quantum-like structures? In other terms: Why  pure quantum measurements, as considered in relation to microphysical systems, appear (so far) to be so effective in modeling the most diverse data obtained in cognitive experiments?

This high degree of universality of the Born rule, associated with pure quantum measurements, is quite surprising, as it is not evident at all that what is usually done in experiments like those conducted  in cognitive science would be equivalent to what physicists do in experiments with microscopic entities. Indeed, each subject participating in a cognitive experiment necessarily brings into it the uniqueness of her/his mind, i.e., the uniqueness of her/his \emph{forma mentis}, with its specific conceptual network forming its inner memory structure. In other terms, it is as if in a physics' laboratory each single outcome was obtained using a different measuring apparatus, every time adjusted in a different way, and possibly working according to different internal principles. 

What we mean to say is that each participant, because of the specificities of her/his mind structure, should be associated with a different statistics of outcomes. This means that in a typical cognitive experiments, involving a number of different subjects, different ``ways of choosing'' (or different ``ways of estimating how participants would choose'') a good exemplar of a given concept are involved, and to each of these different ways of choosing, different probabilities should in principle be associated. This means that a cognitive experiment performed with a number of different subjects should be considered as a collection of different measurements, one for each participant. And each one of  these different measurements should a priori be associated with probabilities having different numerical values (as each participant chooses the outcomes differently), so that the overall probabilities deduced from the results of all the participants in the experiment are in fact  averages over all these different probabilities.

Of course, we are here assuming that the different participants can all be ideally subjected to the same initial condition. This is certainly the case if, for instance, it corresponds to a concept in a well-defined state, as expressed by a word, or a phrase, in relation to a specific question. On the other hand, if the initial condition corresponds to a less precisely defined situation, for instance in a context requiring a decision, then fluctuations in the initial state should also be considered. But these fluctuations can in principle be reduced, whereas those relative to the measurement context, incorporated in each participant's mind, cannot. 

So, a typical cognitive experiment is actually made of different quantum-like measurements, delivering different numerical values for the outcomes (i.e., for the exemplars), but all these different measurements are considered to be unidentifiable, and therefore are not distinguished in the final statistics. This means that the experimenters usually proceed as if they would lack knowledge about the different ``ways of choosing'' employed by the different participants, and simply average over all their individual ``hidden measurements,'' to obtain the final statistics. Considering the above, we are undoubtedly confronted with a little mystery: How is it possible that the very specific Born rule, associated with pure quantum structures, appears to be so good in describing the data gathered from experiments which are in fact statistical mixtures of different measurements, associated with different probabilities?

Here something quite surprising apparently happens: when averaging over different types of measurements, described by probabilities which are in principle different from those obtained by the quantum mechanical Born rule, the result is nevertheless statistically equal to the Born rule. This  means that it should be possible to understand orthodox quantum mechanics as a theory describing the probabilistics of outcomes for measurements that are mixtures of all imaginable types of measurements, and this would also explain why the quantum statistics is so effective, in so many regions of reality, also regarding its numerical statistical predictions.

It is precisely the purpose of the present article, and of its second part \citep{AertsSassolideBianchi2014a}, to show that what is indirectly suggested by the experiments performed by cognitive scientists is actually true, i.e., that orthodox quantum probabilities, described by the Born rule, can be interpreted as the probabilities of a \emph{first-order non-classical theory}, describing situations in which an experimenter doesn't know anything about the nature of the interaction between the entity performing the measurement (in physics, the measuring apparatus; in cognitive sciences, the participating subject) and the entity being measured (in physics, a microscopic ``particle'' in a given state; in cognitive sciences, a concept in a given state).

More precisely, our aim is to introduce and motivate a notion of \emph{universal measurement}, defined as an average over all possible kinds of measurements, and show that such average gives numerically exactly rise to the Born rule of quantum mechanics, thus providing what we think is a fascinating explanation of why the quantum statistics performs so well, in so many experimental ambits. It would do so because it can be understood as a first order theory in the modeling of measurement data. 

It should be mentioned that the idea of universal measurements was firstly introduced by one of us, more than one decade ago \citep{Aerts1998, Aerts1999b}, in the ambit of his analysis of classical, quantum and intermediary structures. It was already suggested at that time that when we are in a condition of maximal lack of knowledge, in a given experimental situation, what we may actually end up performing is a ``huge'' kind of measurement -- a universal measurement -- consisting in choosing at-random between all possible measurements. The interesting idea that was already brought forward then, although only as a conjecture, is that if there is ``one'' physical reality, then there should also be ``one'' universal measurement, connecting an initial state to a final state. 

This uniqueness of universal measurements was indirectly suggested by the existence of a famous theorem of quantum mechanics, \emph{Gleason's theorem}, which affirms that ``if the transition probability depends only on the state before the measurement and on the eigenstate of the measurement that is actualized after the measurement, then this transition probability is equal to the quantum transition probability.'' And since this Gleason property (dependence of the transition probability only on the state before the measurement and the eigenstate that is actualized after the measurement) is exactly a property that is satisfied, by definition, by a universal measurement, the theorem suggested (within the limit of Hilbertian structures) the possibility that the transition probabilities connected with universal measurements would be precisely the quantum mechanical transition probabilities, described by the Born rule.

The idea, however, remained only conjectural at that time, because of difficulties related to the so-called \emph{Bertrand Paradox}, i.e., to the fact that probabilities may depend on the randomization method  chosen to perform a uniform average, when the number of possible cases is infinite \citep{Bertrand1889}. But these difficulties have now been overcome, thanks to a transparent definition of the uniform randomization over measurements, conceived  as a limit of randomized  discrete systems, much in the spirit of what is traditionally done in physics when averages over paths are performed, for instance in the study of Brownian motion. 

The above was to emphasize that the results contained in this paper, and in \citet{AertsSassolideBianchi2014a}, are of interest not only for cognitive scientists, but also for physicists, as the possibility of understanding the measurements of quantum mechanics as universal measurements is new and relevant in both domains of investigation. Now, to be able to study the ``huge'' average associated with a universal measurement, we need a suitable and sufficiently general theoretical framework, and this brings us to the second element of novelty contained in this paper. Indeed, this framework will be provided by an idealized system -- that we call the \emph{general tension-reduction} (GTR) model -- able to describe a virtually infinite number of different measurement situations, in different dimensions, ranging from the classical deterministic ones, associated with processes of pure \emph{discovery}, to the ``solipsistic'' indeterministic ones, associated with processes of pure \emph{creation}, with in the middle the pure quantum regime, expressing a sort of equilibrium between these two extreme conditions of pure discovery and pure creation. 

It should be mentioned that the GTR-model that we introduce here  is a non-trivial multidimensional generalization of what is known as the \emph{sphere-model} \citep{Aertsetal1997b}, which in turn is a generalization of the so-called \emph{$\epsilon$-model} \citep{Aerts1998, Aerts1999b, SassolideBianchi2013b} and, interestingly, for the special cases of 1, 2 and 3 dimensions (i.e., 2, 3 and 4 outcomes, respectively), it describes a system which in principle could be realized in a laboratory, by means of specially designed materials. One of the great advantages of the model is that it allows to fully understanding and even visualizing what goes on during a measurement, when the state of the entity under investigation collapses into an eigenstate of the measured observable, thus providing a considerable insight into many aspects of quantum structures. In particular, it explicitly shows that quantum and quantum-like measurements can be understood in terms of hidden (potential) measurement interactions, which are actualized in an unpredictable way each time an experiment is performed, in accordance with the so-called \emph{hidden-measurement approach} \citep{Aerts1986, Aerts1998, Aerts1999b, Coecke1995, SassolideBianchi2013a}.

Another important advantage of the GTR-model is that it allows to represent all kinds of possible distributions of hidden interactions, of which the pure quantum one only constitutes a very special case, corresponding to the choice of a uniform probability density $\rho_u$. Thanks to this great level of generality, it becomes possible to use the model as a general theoretical framework to state and derive our result regarding the correspondence between universal measurements and quantum measurements (the result will only be enunciated and explained in the present article, the  proof being given in \citet{AertsSassolideBianchi2014a}). Also, considering that the GTR-model is a more general theoretical framework than the Hilbert-model, it can be exploited to model situations where the first order approximation expressed by the Born rule would not be sufficient, i.e., when some knowledge about the fluctuations present in the experimental context would be available, a possibility which is more likely to manifest in cognitive experiments than in physics experiments.  

The work is organized as follows. In Sec.~\ref{Quantum Probabilities for a Single Observable}, we present some basic elements of the quantum formalism, to define notations and to allow  to more easily establish, in the subsequent sections,  the correspondence between quantum measurements and the measurement described in the GTR-model. In Sec.~\ref{Thepolytopemodel},  we start by analyzing the GTR-model in the special case where the probability density is uniform $\rho \equiv \rho_u$. This special case, which we refer to as the \emph{uniform tension-reduction} (UTR) model, is already sufficiently general to describe all possible probabilities arising in a single measurement, a fact that will be emphasized in Sec.~\ref{RepresentingProbabilities}, by means of a representation theorem. 

In our study of the UTR-model (the GTR-model with a uniform probability density), we will proceed in a pedestrian way, by first describing the one-dimensional (two-outcome) and two-dimensional (three-outcomes) situations,  then generalizing the description to an arbitrary number of dimensions. Although the model is per se  abstract, in our analysis we will mostly concentrate on one of its possible physical realizations, using special elastic hypermembranes which can break and collapse in a specific way. 

In Sec.~\ref{Entanglementrhomodel}, we use the UTR-model to shed some light into the phenomenon of entanglement, showing that the process of \emph{emergence} it subtends requires additional dimensions to be described. In Sec.~\ref{RepresentingProbabilities}, as we said, we state a general representation theorem, showing that the UTR-model (and equally so the Hilbert-model) is a ``universal probabilistic machine,'' able to represent any possible probabilities emerging from a single measurement. We also show that these probabilities can either be understood as the result of the presence of a uniform mixture of pure measurements or, in a sort of complementary picture, of a uniform mixture of initial states. 

In Sec.~\ref{Non-uniform}, we introduce the more general GTR-model, which also admits non-uniform probability densities (thus generalizing the UTR-model), and use it to motivate a notion of \emph{universal measurement}, which will be defined in a physically transparent  and mathematically precise way. This will allow us to state our theorem about the equivalence between universal measurements and measurements characterized by uniform fluctuations (and therefore their correspondence with the quantum mechanical Born rule), which  will be formally proven only in the second part of this article \citep{AertsSassolideBianchi2014a}. 

In Sec.~\ref{nonhilbert}, we  analyze in some detail the ampler structural richness of the GTR-model, by investigating, in the two-outcome case, the probabilities associated with \emph{sequential measurements}, showing that they cannot in general be fitted into a Hilbertian or Kolmogorovian structure. Finally, in Sec.~\ref{Conclusion}, we offer some conclusive remarks.

\vspace{-0.4cm}
\section{Quantum Probabilities for a Single Observable}
\label{Quantum Probabilities for a Single Observable}
\vspace{-0.4cm}

In this section we present some elements of the basic formalism of quantum mechanics, in relation to the measurement of a finite dimensional observable, which can either be degenerate or non-degenerate. In doing so, we will also describe the special case of a compound system made of two entities, and emphasize the difference between product and non-product (entangled) states. 

In orthodox quantum theory, the state of an entity (for a physicist it can be a microscopic entity, such as an electron, for a cognitive scientist, a concept, or a situation apt for a decision process)  is described by a  vector space over the field ${\mathbb C}$ of complex numbers -- the so-called Hilbert space ${\mathcal H}$ -- equipped with a (sesquilinear) inner product $\langle \cdot |  \cdot \rangle$, which maps two vectors $|\phi\rangle$, $|\psi\rangle$ to a complex number $\langle \phi|\psi\rangle$, and consequently with a norm $\| |\psi\rangle \| \equiv \sqrt{\langle \psi|\psi\rangle}$, which assigns a positive length to each vector. In this article we only consider  Hilbert spaces having a finite number of dimensions, and will denote  $\mathcal{H}_{N}$ a Hilbert space which is an $N$-dimensional vector space.

An observable is a measurable quantity of the entity under consideration, and in quantum theory is represented by a self-adjoint operator $A$, acting on vectors of the Hilbert space, i.e., $A: |\psi\rangle \rightarrow A|\psi\rangle$. In our case, being the Hilbert space $N$-dimensional,  $A$  can be entirely described by means of its $N$ eigenvectors $|a_i\rangle$ and the associated (real) eigenvalues $a_i$, obeying the eigenvalue relations $A|a_i\rangle = a_i |a_i\rangle$, for all $i\in \{1,\dots,N\}\equiv I_{N}$. If the eigenvectors have been duly normalized, so that in addition to the orthogonality relation $\langle a_i|a_j\rangle = \delta_{ij}$, $i,j\in I_{N}$, they also obey the completeness relation $\sum_{i\in I_{N}} |a_i\rangle\langle a_i| = \mathbb{I}$, where $\mathbb{I}$ denotes the unit operator, they can be used to construct the orthogonal projections $P_i\equiv |a_i\rangle\langle a_i|$, $i\in I_{N}$, obeying $\sum_{i\in I_{N}} P_i =\mathbb{I}$, $P_iP_j =P_i\delta_{ij}$, $i,j\in I_{N}$, which in turn can be used to write the observable $A$ as the (spectral) sum:
\begin{eqnarray}
\label{observable-hilbert}
A = \mathbb{I} A = \left[\sum_{i\in I_{N}} P_i\right]A=\sum_{i\in I_{N}} a_i P_i.
\end{eqnarray}

Similarly, if $|\psi\rangle\in \mathcal{H}_{N}$, $\|\psi\|^2=\langle \psi|\psi\rangle=1$, is a normalized vector describing the state of the entity, it can be written as the sum: 
\begin{eqnarray}
\label{state-hilbert}
|\psi\rangle = \mathbb{I} |\psi\rangle = \left[\sum_{i\in I_{N}} P_i\right]|\psi\rangle =\sum_{i\in I_{N}} |a_i\rangle\langle a_i|\psi\rangle = \sum_{i\in I_{N}} \sqrt{x_i}e^{i\alpha_i} |a_i\rangle,
\end{eqnarray}
where for the last equality we have written the complex numbers $\langle a_i|\psi\rangle$ in the polar form $\langle a_i|\psi\rangle = \sqrt{x_i}e^{i\alpha_i}$. Clearly, being $|\psi\rangle$ normalized to $1$, the positive real numbers $x_i$ must obey:
\begin{eqnarray}
\label{sumofthexi}
\sum_{i\in I_{N}}x_i = 1.
\end{eqnarray}
\\
\emph{The non-degenerate case}
\\
When we measure an observable $A$ in a practical experiment (a physicist does so by letting the microscopic entity interact with a macroscopic measuring apparatus, a psychologist by letting a human concept interact with a human mind, according to a certain protocol, if concepts are studied, or by collecting the decision results, if situations lending themselves to human decisions are studied), we can obtain one of the $N$  eigenvalues $a_i$, $i\in I_{N}$, and if  these $N$ eigenvalues are all different, we say that the spectrum of $A$ is \emph{non-degenerate}. Consequently, the measurement has $N$ distinguishable possible outcomes. 

In general terms, the measurement of an observable $A$ is a process during which the state of the entity undergoes an abrupt transition -- called ``collapse'' in the quantum jargon -- passing from the initial state $|\psi\rangle$ to a final state which is one of the eigenvectors $|a_i\rangle$ of $A$, associated with the eigenvalue $a_i$, $i\in I_{N}$. The process is non-deterministic, and we can only describe it in probabilistic terms, by means of a ``golden rule,'' called the \emph{Born rule}, which states the following: the probability $P(|\psi\rangle\to|a_i\rangle)$, for the transition $|\psi\rangle\to|a_i\rangle$, is given by the  square of the length of the vector $P_i|\psi\rangle$, i.e., the square of the length  of the initial vector once it has been projected onto the eigenspace of $A$ corresponding to the eigenvalue $a_i$. More explicitly: 
\begin{eqnarray}
\label{Born}
\lefteqn{P(|\psi\rangle\to|a_i\rangle) = \| P_i|\psi\rangle\|^2 = \langle\psi|P_i P_i|\psi\rangle}\nonumber\\
&= \langle\psi| P_i|\psi\rangle = \langle\psi|a_i\rangle\langle a_i|\psi\rangle =|\langle a_i|\psi\rangle|^2 = x_i,
\end{eqnarray}
for all $i\in I_{N}$. And of course, according to (\ref{sumofthexi}), we have: 
\begin{eqnarray}
\label{totalprob=1}
\sum_{i\in I_{N}} P(|\psi\rangle\to|a_i\rangle) = \sum_{i\in I_{N}}x_i = 1,
\end{eqnarray}
as it must be, by definition of a probability. 
\\
\vspace{-0.3cm}
\\
\emph{The degenerate case}
\\
We  consider now a \emph{degenerate} observable $A$. This means that some of the $a_i$ will have the same value, and therefore are not distinguishable, as outcomes, by the experimenter. To describe this situation, we consider $n$ disjoint subsets $I_{M_k}$ of $I_{N}\equiv \{1,\dots,N\}$,  $k=1\dots,n$,  having $M_k$ elements each, with $0\leq M_k\leq N$ and $\sum_{k=1}^{n} M_k=N$, so that $\cup_{k=1}^{n} I_{M_k} =I_{N}$. We then assume that the eigenvectors $|a_i\rangle$ whose index belong to a same set $I_{M_k}$ are all associated with a same eigenvalue $a_{I_{M_k}}$,  $M_k$ times degenerate. Therefore, defining the projectors $P_{I_{M_k}}\equiv \sum_{i\in I_{M_k}} P_i$, onto the $M_k$-dimensional eigenspace associated with the eigenvalues  $a_{I_{M_k}}$, (\ref{observable-hilbert}) becomes:
\begin{eqnarray}
\label{observable-hilbert-degenerate}
A = \sum_{k=1}^{n} \sum_{i\in I_{M_k}} a_i P_i =   \sum_{k=1}^{n} a_{I_{M_k}} \left[\sum_{i\in I_{M_k}} P_i\right] = \sum_{k=1}^{n} a_{I_{M_k}} P_{I_{M_k}},
\end{eqnarray}
and of course, (\ref{observable-hilbert-degenerate}) gives back (\ref{observable-hilbert}) when each of the sets $I_{M_k}$ is a singleton $\{k\}$, i.e., a set containing the single element $k$, and consequently $n=N$.

For a degenerate observable we cannot anymore associate one-dimensional eigenspaces to the distinguishable eigenvalues. Accordingly, measurements will now produce state transitions of the form $|\psi\rangle \to |\psi_{I_{M_k}}\rangle$, $k=1,\dots,n$, where: 
\begin{eqnarray}
\label{new-state}
|\psi_{I_{M_k}}\rangle &=& {P_{I_{M_k}} |\psi\rangle \over \| P_{I_{M_k}} |\psi\rangle\|}=   {\sum_{i\in I_{M_k}} P_i|\psi\rangle\over \sqrt{\sum_{i\in I_{M_k}}\langle\psi|P_i|\psi\rangle}}={\sum_{i\in I_{M_k}} \sqrt{x_i}e^{i\alpha_i} |a_i\rangle\over \sqrt{\sum_{i\in I_{M_k}}x_i}}\nonumber\\
&=&\sum_{i\in I_{M_k}} \sqrt{x_i\over \sum_{j\in I_{M_k}}x_j }\,\,e^{i\alpha_i} |a_i\rangle.
\end{eqnarray}
According to the Born rule, for $k=1,\dots,n$, we  have the transition probabilities:
\begin{eqnarray}
\label{Born-degenerate}
\lefteqn{P(|\psi\rangle\to|\psi_{I_{M_k}}\rangle) = \| P_{I_{M_k}}|\psi\rangle\|^2 = \langle\psi|P_{I_{M_k}} P_{I_{M_k}}|\psi\rangle = \langle\psi|P_{I_{M_k}}|\psi\rangle}\nonumber\\
&=\sum_{i\in I_{M_k}} \langle\psi|a_i\rangle\langle a_i|\psi\rangle =\sum_{i\in I_{M_k}} |\langle a_i|\psi\rangle|^2 = \sum_{i\in I_{M_k}} x_i,
\end{eqnarray}
and of course
\begin{eqnarray}
\label{totalprob=1-degenerate}
\sum_{k=1}^{n}   P(|\psi\rangle\to|\psi_{I_{M_k}}\rangle) = \sum_{k=1}^{n} \sum_{i\in I_{M_k}} x_i =  \sum_{i\in I_{N}}x_i = 1.
\end{eqnarray}
\\
\emph{Compound systems}
\\
To illustrate the importance of the above distinction between degenerate and non-degenerate observables, we describe the important case of compound systems, consisting in more than a single entity. For sake of simplicity, we limit our discussion to the case of a compound system made of only two entities, which can only be in two different states. Typically, a physicist will consider two spin-${1\over 2}$ entities, like two electrons, whereas a psychologist studying concepts will consider the combinations of two concepts,  allowing for each of them only two possible exemplars. Then, the Hilbert space is the $4$-dimensional complex space $\mathcal{H}_4=\mathbb{C}^4$, and since there are two entities, it can also be described as the tensor product $\mathcal{H}_2\otimes \mathcal{H}_2 = \mathbb{C}^2\otimes\mathbb{C}^2$, where the first $2$-dimensional Hilbert space $\mathcal{H}_2 = \mathbb{C}^2$ is associated with the first entity (indicated by the index $1$ in the following), and the second identical Hilbert space is associated with the second entity (indicated by the index $2$).
 
A product observable $A$ can then be written as the tensor product  $A= A_1\otimes A_2$, where  $A_1$ acts on the first entity and $A_2$ acts on the second one. More specifically, $A_1\otimes A_2$ corresponds to the coincident measurement of $A_1$ on the first entity and $A_2$ on the second one. This can  be also expressed by writing  $A$ as the ordinary product of two commuting observables: $A=(A_1\otimes \mathbb{I})(\mathbb{I}\otimes A_2)=(\mathbb{I}\otimes A_2)(A_1\otimes \mathbb{I})$, where $A_1\otimes \mathbb{I}$ is the observable acting on the first entity via $A_1$, but doing nothing to the second entity, whereas $\mathbb{I}\otimes A_2$ does nothing to the first entity, but acts on the second one via $A_2$. In other terms, $A_1\otimes \mathbb{I}$ corresponds to an observation only on the first entity, whereas $\mathbb{I}\otimes A_2$ to an observation only on the second one, and their product corresponds to a coincident observation on  
the entity which is a compound of both entities.

Let us show that $A= A_1\otimes A_2$ has a non-degenerate spectrum of eigenvalues, whereas $A_1\otimes \mathbb{I}$ and $\mathbb{I}\otimes A_2$ have doubly degenerate eigenvalues. For this, we introduce in $\mathcal{H}_4$ the four (tensor product) base vectors $|\mu,\nu\rangle \equiv |\mu\rangle_1 \otimes |\nu\rangle_2$, $\mu,\nu\in\{1,2\}$, where $|\mu\rangle_1$, $\mu=1,2$, are the two eigenvectors of $A_1$, with non-degenerate eigenvalues $a_{1;\mu}$, $\mu=1,2$, and  $|\nu\rangle_1$, $\nu=1,2$, are the two eigenvectors of $A_2$, with non-degenerate eigenvalue $a_{2;\nu}$, $\nu=1,2$. Clearly, $A|\mu,\nu\rangle= A_1\otimes A_2|\mu\rangle_1\otimes |\nu\rangle_2=A_1|\mu\rangle_1\otimes A_2|\nu\rangle_2 = a_{1;\mu}|\mu\rangle_1 \otimes a_{2;\nu}|\nu\rangle_2=a_{1;\mu} a_{2;\nu}|\mu,\nu\rangle$, that is, the $|\mu,\nu\rangle$ are the eigenvectors of $A$, associated with the four  eigenvalues $a_{1;\mu} a_{2;\nu}$, $\mu,\nu=1,2$, which are all distinct, and so $A$ is non-degenerate. On the other hand, $A_1\otimes \mathbb{I}|\mu,\nu\rangle=A_1\otimes \mathbb{I}||\mu\rangle_1\otimes |\nu\rangle_2=A_1|\mu\rangle_1\otimes \mathbb{I}||\nu\rangle_2 = a_{1;\mu}|\mu\rangle_1 \otimes 1|\nu\rangle_2=a_{1;\mu} |\mu,\nu\rangle$, that is, the $|\mu,\nu\rangle$ are also eigenvectors of $A_1\otimes \mathbb{I}$, but this time both $|\mu,1\rangle$ and $|\mu,2\rangle$ are associated with the same eigenvalue $a_{1;\mu}$, which therefore is doubly degenerate, for each $\mu=1,2$. And of course, the same holds true for $\mathbb{I}\otimes A_2$.

In general, a state $|\psi\rangle$ of the two-entity system can be written in the above eigen-basis as the superposition: 
\begin{eqnarray}
\label{generalstate}
|\psi\rangle = \sqrt{x_1} e^{i\alpha_1}|1,1\rangle +  \sqrt{x_2} e^{i\alpha_2}|1,2\rangle +\sqrt{x_3} e^{i\alpha_3}|2,1\rangle+\sqrt{x_4} e^{i\alpha_4}|2,2\rangle,
\end{eqnarray}  
and according to the above, a measurement of $A$ can produce four different outcomes, associated with the probabilities: 
\begin{eqnarray}
&&P(|\psi\rangle\to |1,1\rangle) = |\langle 1,1|\psi\rangle|^2 = x_1,\\ 
&&P(|\psi\rangle\to |1,2\rangle) = |\langle 1,2|\psi\rangle|^2 =x_2,\\
&&P(|\psi\rangle\to |2,1\rangle) = |\langle 2,1|\psi\rangle|^2 = x_3,\\ 
&&P(|\psi\rangle\to |2,2\rangle) = |\langle 2,2|\psi\rangle|^2 =x_4.
\end{eqnarray}  
On the other hand, a measurement of $A_1\otimes \mathbb{I}$ can only produce two different outcomes, associated with the probabilities: 
\begin{eqnarray}
&&P(|\psi\rangle\to |\psi_{\{1,2\}}\rangle) = \sum_{\nu =1}^2 |\langle 1,\nu|\psi\rangle|^2 = x_1 + x_2,\\
&&P(|\psi\rangle\to |\psi_{\{3,4\}}\rangle) = \sum_{\nu =1}^2 |\langle 2,\nu|\psi\rangle|^2 = x_3 + x_4,
\end{eqnarray}  
and similarly, a  measurement of $\mathbb{I}\otimes A_2$ is associated with the two transition probabilities: 
\begin{eqnarray}
&&P(|\psi\rangle\to |\psi_{\{1,3\}}\rangle) =  \sum_{\mu =1}^2 |\langle \mu,1|\psi\rangle|^2 = x_1 + x_3,\\
&&P(|\psi\rangle\to |\psi_{\{2,4\}}\rangle) =  \sum_{\mu =1}^2 |\langle \mu,2|\psi\rangle|^2 = x_2 + x_4.
\end{eqnarray}    

We conclude this brief description of the two-entity system by mentioning the difference between \emph{product} (non-entangled) states, and \emph{non-product} (entangled) states. The latters are associated with the possibility of creating correlations when measurements are jointly performed on the entity which is the compound of both entities, whereas the formers cannot produce any correlations. A product state is a state of the form $|\psi\rangle_1\otimes |\psi\rangle_2$, where $|\psi\rangle_1= \sqrt{a} e^{i\alpha}|1\rangle_1 +  \sqrt{b} e^{i\beta}|2\rangle_1$, with $a + b =1$, and $|\psi\rangle_2= \sqrt{c} e^{i\delta}|1\rangle_2 +  \sqrt{d} e^{i\gamma}|2\rangle_2$, with $c + d =1$. This means that: 
\begin{eqnarray}
\label{productstate}
\lefteqn{|\psi\rangle_1\otimes |\psi\rangle_2 = \sqrt{ac}e^{i(\alpha +\delta)}|1,1\rangle + \sqrt{ad} e^{i(\alpha +\gamma)}|1,2\rangle}\nonumber\\
&+\,\sqrt{bc} e^{i(\beta +\delta)}|2,1\rangle+\sqrt{bd} e^{i(\beta +\gamma)}|2,2\rangle.
\end{eqnarray} 
Therefore, if we use a product state (\ref{productstate}), instead of a general state (\ref{generalstate}), we obtain for the different transition probabilities associated with the measurements of the three observables $A=A_1\otimes A_2$, $A_1\otimes \mathbb{I}$ and  $\mathbb{I}\otimes A_2$:
\begin{eqnarray}
&&P(|\psi\rangle\to |1,1\rangle) = ac,\quad P(|\psi\rangle\to |1,2\rangle) = ad,\\
&&P(|\psi\rangle\to |2,1\rangle) =  bc,\quad P(|\psi\rangle\to |2,2\rangle)  = bd,\\
&&P(|\psi\rangle\to |\psi_{\{1,2\}}\rangle) = a(c+d)=a,\\ 
&&P(|\psi\rangle\to |\psi_{\{3,4\}}\rangle) = b(c+d) = b,\\
&&P(|\psi\rangle\to |\psi_{\{1,3\}}\rangle) = c(a+b)=c,\\
&&P(|\psi\rangle\to |\psi_{\{2,4\}}\rangle)= d(a+b) =d.
\end{eqnarray} 

Setting $x_1=ac$, $x_2=ad$, $x_3=bc$, and $x_4=bd$, we thus obtain that state (\ref{generalstate}), to be a product state, must obey
\begin{eqnarray}
&&x_1=(x_1+x_2)(x_1+x_3),\quad x_2=(x_1+x_2)(x_2+x_4),\nonumber\\
&&x_3= (x_3+x_4)(x_1+x_3),\quad x_4=(x_3+x_4)(x_2+x_4)\label{productrelations}.
\end{eqnarray} 
Clearly, the eigenvectors $|\mu,\nu\rangle \equiv |\mu\rangle_1 \otimes |\nu\rangle_2$, $\mu,\nu=1,2$, trivially obey the above relations. For instance, state $|1,1\rangle$ corresponds to  $x_1=1$ and $x_2=x_3=x_4=0$, which obviously obey (\ref{productrelations}). On the other hand, a typical entangled state, like a so-called singlet state of the form $|\psi\rangle = {1\over\sqrt{2}}(|1,2\rangle - |2,1\rangle)$, corresponding to $x_2=x_3 ={1\over 2}$ and $x_1 =x_4=0$,   disobeys (\ref{productrelations}), as by replacing these values in the first relation above, we obtain $0={1\over 4}$, which is a contradiction.

\vspace{-0.4cm}
\section{The UTR-model}
\label{Thepolytopemodel}
\vspace{-0.4cm}

In this section we present an abstract model, which we call the \emph{uniform tension-reduction} (UTR) model, allowing for the description and representation of general -- single observable -- measurement situations, characterized by an arbitrary (finite) number of different possible outcomes. The model is universal, in the sense that probabilities of any numerical value can be represented by it. Being an abstract construction, its usefulness does not depend on the existence of possible physical realizations of its structure. However, the possibility of describing the UTR-model by means of a specific mechanical realization of it, will prove to be quite helpful in gaining greater intuition about the hidden structure which is ``behind'' a measurement in general, and a quantum measurement in particular. This is what we shall do below, keeping always in mind that the ``universal machine''  we shall describe only constitutes one of many possible physical realizations of the abstract structure of the UTR-model (for a different physical realization, in the two-outcome case, see for instance \citep{Aerts1986}). 

It is important to clarify what we mean exactly when we say that the UTR-model admits a mechanical realization. By this, we mean that we can define a mechanical system, i.e., a ``machine,'' functioning in a logical way, able to represent the outcomes of whatever measurement, and the associated probabilities. But by this we don't necessarily mean that such system can be constructed in reality, using today known materials, and this for at least three reasons: the first one is that a theoretical, abstract model is always  an idealization of more concrete systems, which can only constitute approximations (in the same way that real Newtonian systems approximate idealized Newtonian systems, when for instance they assume that there are no frictions); the second one is that we are certainly not able today to manufacture materials that behave exactly in the same way the model behaves, although we may be able to do so in a near future; the third one is that in any case, for more than four outcomes, the machine necessitates more than three spatial dimensions in order to operate, and of course we cannot construct macroscopic objects of four or more spatial dimensions. 

More specifically, we will realize the different logical operations in our UTR-model by using the key notion of a `breakable elastic membrane' (which will become a hypermembrane for more than 4 outcomes). This because, to the best of our knowledge, it provides the simplest physical realization of a measurement situation allowing not only for a clear identification of the source of indeterminism, but also for the possibility of using the Lebesgue measure to derive a corresponding probability calculus. In addition to that, the membranes, with their `tension lines', are very close to what we imagine to possibly be the dynamics unfolding in our brains, during a decision making process. In other terms, the UTR-model is perfectly compatible with what our intuition tells us, regarding the way we create and break a set of competing tensions in our mind, in a given decisional context, as we will better explain in the last part of this section. 

But before that, we have to proceed step by step, describing first the two-outcome situation ($N=2$), then the three-outcome situation ($N=3$), and finally the general situation, with an arbitrary number $N$ of outcomes.
\\
\vspace{-0.3cm}
\\
\emph{The $N=2$ case, with two outcomes}
\\
The entity is a simple material point particle (to be understood also as an abstract entity representative of the state of the system under consideration)
living in a Euclidean space ${\mathbb R}^{n}$, $n\geq 2$, and measurements, which will be denoted $e_{\{1\}\{2\}}$, can only have two outcomes. 
The procedure to follow to perform $e_{\{1\}\{2\}}$ is the following. The experimenter takes a sticky, breakable and uniform elastic band, and stretches it over a $1$-dimensional simplex\footnote{A simplex is a generalization of the notion of a triangle. A 1-simplex is a line segment; a 2-simplex is an equilateral triangle; a 3-simplex is a tetrahedron; a 4-simplex is a pentachoron; and so on.} $S_1$ (1-simplex), generated by two orthonormal vectors $\hat{\bf{x}}_1$ and $\hat{\bf{x}}_{2}$. Once the uniform elastic band is in place, the particle, by moving \emph{deterministically} towards it, along a trajectory that will depend on the structure of the state space, sticks to it at a particular point ${\bf{x}}=x_1 \hat{\bf{x}}_1 + x_2 \hat{\bf{x}}_2$, $x_1 + x_2 = 1$, defining the state of the particle on the elastic (we represent Euclidean vectors in bold).

For instance, if the point particle is representative of a two-state quantum-mechanical system (qubit), the state space can be put into correspondence with the \emph{Bloch sphere} (also called the \emph{Poincar\'e sphere}), and the deterministic map that brings the particle in contact with the sticky elastic band corresponds to a movement along a rectilinear path, orthogonal to the latter, as it will be better explained in Sec.~\ref{nonhilbert}.

\begin{figure}[!ht]
\centering
\includegraphics[scale =0.55]{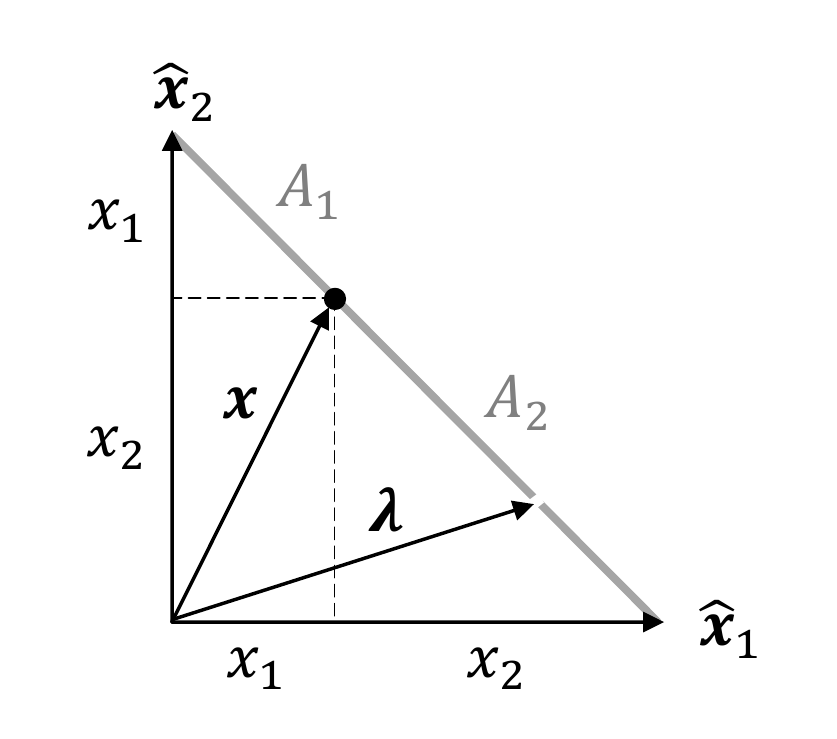}
\caption{The $1$-dimensional (1-simplex) elastic structure attached to the two unit vectors $\hat{\bf{x}}_1$ and $\hat{\bf{x}}_2$, with the two  regions $A_1$ and $A_2$ generated by the presence of the point particle in $\bf{x}$. The vector \mbox{\boldmath$\lambda$}, here in region $A_2$, indicates the point where the elastic breaks. 
\label{1-dimensions}}
\end{figure}

\begin{figure}[!ht]
\centering
\includegraphics[scale =.55]{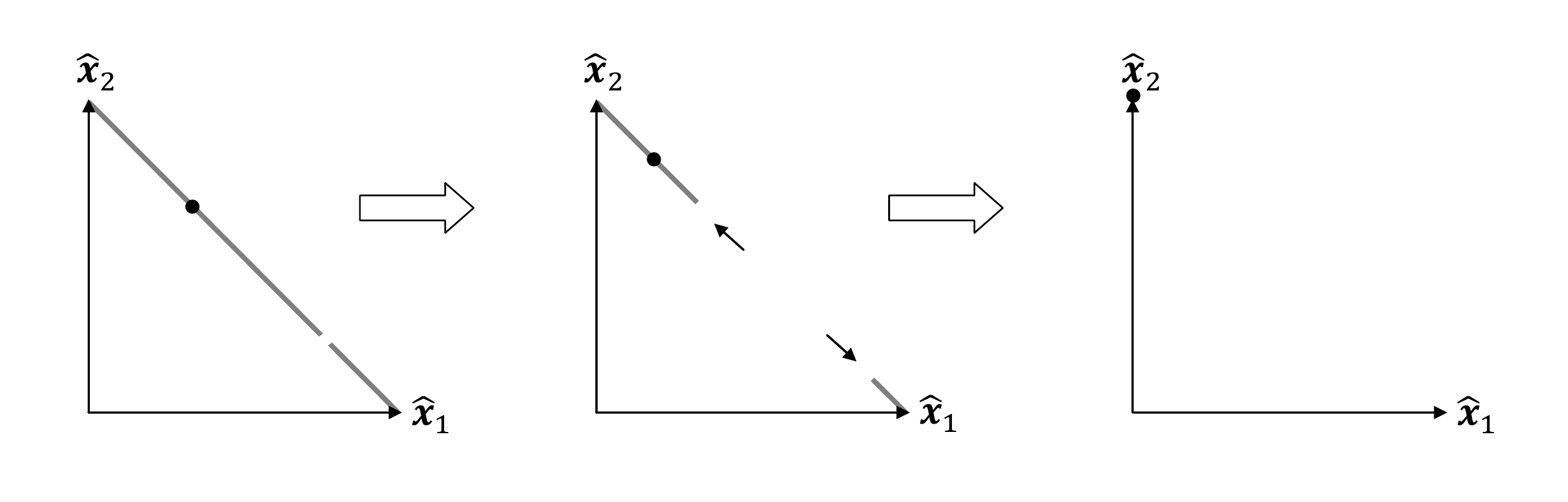}
\caption{The breaking of the elastic causes the particle to be drawn to  point $\hat{\bf{x}}_2$.  
\label{1-dimensionsbreaking}}
\end{figure}

When the particle is in place on the elastic, two disjoint regions $A_1$ and $A_2$ can be distinguished, which are respectively the region bounded by vectors $\hat{\bf{x}}_2$ and $\bf x$, and the region bounded by vectors $\bf x$ and $\hat{\bf{x}}_1$ (see Fig.~\ref{1-dimensions}).
Then, after some time, as the uniform elastic band is made of a breakable material, it inevitably breaks, at some a priori unpredictable point \mbox{\boldmath$\lambda$} (see Fig.~\ref{1-dimensions}). If $\mbox{\boldmath$\lambda$}\in A_1$,  the elastic band, when it contracts, it  draws the particle to point $\hat{\bf{x}}_1$, whereas if $\mbox{\boldmath$\lambda$}\in A_2$, the elastic  draws the particle to point $\hat{\bf{x}}_2$, which is the ``collapse process'' depicted in Fig.~\ref{1-dimensionsbreaking}.

We can observe that to each breaking point \mbox{\boldmath$\lambda$}, it corresponds a specific interaction between the particle and the elastic band, which draws the former  to its final state $\hat{\bf{x}}_1$, or $\hat{\bf{x}}_2$ (the two possible outcomes of the measurement). In other terms, the measurement $e_{\{1\}\{2\}}$ is a collection of hidden (potential) \emph{pure measurements}, only one of which is each time selected (actualized), when the elastic breaks. Let us observe that given the particle state ${\bf x}$, all pure measurements but one are deterministic, as for $\mbox{\boldmath$\lambda$} ={\bf x}$ the outcome  remains clearly indeterminate, in the classical sense of a system in a condition of unstable equilibrium.  

To calculate the probabilities of the two outcomes, one needs to observe that being the elastic uniform, all its points have exactly the same probability to break (the elastic is a physical realization of a uniform probability density). Therefore, the probability $P(A_i)$ for the elastic to break in  region $A_i$, $i=1,2$, is simply given by the ratio between the length of the segment $A_i$ (the Lebesgue measure $\mu_L(A_i)$ of region $A_i$) and the total length $\|\hat{\bf{x}}_{2}-\hat{\bf{x}}_{1}\| = \sqrt{2}$ of the band (the the Lebesgue measure $\mu_L(S_1)$ of the 1-simplex): $P(A_i) = {\mu_L(A_i)\over \mu_L(S_1)}={\mu_L(A_i)\over\sqrt{2}}$. From Pythagorean theorem, and $x_1 + x_2 = 1$, it immediately follows that (see Fig.~\ref{1-dimensions}) 
$\mu_L(A_i)=\sqrt{x_1^2 + x_1^2}=\sqrt{2}x_1$, so that $P(A_i)= x_i$, $i=1,2$. And since the particle is drawn to  $\hat{\bf{x}}_{i}$ when the elastic breaks in $A_i$, the probability $P({\bf{x}}\to \hat{\bf{x}}_{i})$ for the transition ${\bf{x}}\to \hat{\bf{x}}_{i}$ is precisely the probability $P(A_i)$ for the elastic to break in $A_i$, so that we can write: 
\begin{eqnarray}
\label{x1integral-z-u}
P({\bf{x}}\to \hat{\bf{x}}_{i})= P(A_i) = {\mu_L(A_i)\over \mu_L(S_1)} = x_i, \quad i=1,2.
\end{eqnarray}

In other terms, in accordance with (\ref{Born}), measurement $e_{\{1\}\{2\}}$ is  isomorphic to the measurement of an observable $A$ in a two-dimensional complex Hilbert space ${\mathcal H}_2$, if we  represent the quantum state vector $|\psi\rangle = \sqrt{x_1}e^{i\alpha_1}|a_1\rangle +\sqrt{x_2}e^{i\alpha_2}|a_2\rangle \in {\mathcal H}_2$, by a vector ${\bf{x}} = x_1 \hat{\bf{x}}_1 + x_2 \hat{\bf{x}}_2$, whose components are precisely the transition probabilities (see Sec.~\ref{Quantum Probabilities for a Single Observable}).
\\
\vspace{-0.3cm}
\\
\emph{The $N=3$ case, with three outcomes}
\\
We consider now the slightly more complex situation consisting of measurements which can have three possible outcomes. The entity is always a material point particle,  living in a Euclidean space ${\mathbb R}^{n}$, $n\geq 3$. Different typologies of (non-trivial) measurements can be carried out in this case.  More precisely, we can distinguish four different typologies of measurements: $e_{\{1\}\{2\}\{3\}}$, $e_{\{1,2\}\{3\}}$, $e_{\{1,3\}\{2\}}$ and $e_{\{2,3\}\{1\}}$. We start describing the first one, which corresponds to the situation where all three outcomes can be distinguished by the experimenter (non-degenerate measurement). 

The procedure to follow to perform $e_{\{1\}\{2\}\{3\}}$ is the following. The experimenter takes a sticky, uniformly breakable  elastic membrane and stretches it over a $2$-dimensional simplex $S_2$ generated by  three orthonormal  vectors $\hat{\bf{x}}_1$, $\hat{\bf{x}}_2$ and $\hat{\bf{x}}_{3}$, attaching it to its three vertex points. Once the uniform elastic membrane is in place, the particle, by moving deterministically towards it (along a trajectory that is not important here to specify, which will depend on the structure of the state space; see the discussion at the end of Sec.~\ref{nonhilbert}, for the case of a Hilbertian state space), sticks to it at a particular point ${\bf{x}}=x_1 \hat{\bf{x}}_1 + x_2 \hat{\bf{x}}_2+ x_3 \hat{\bf{x}}_3$, with $x_1 + x_2 + x_3 = 1$, defining the state of the particle on the membrane.

\begin{figure}[!ht]
\centering
\includegraphics[scale =.75]{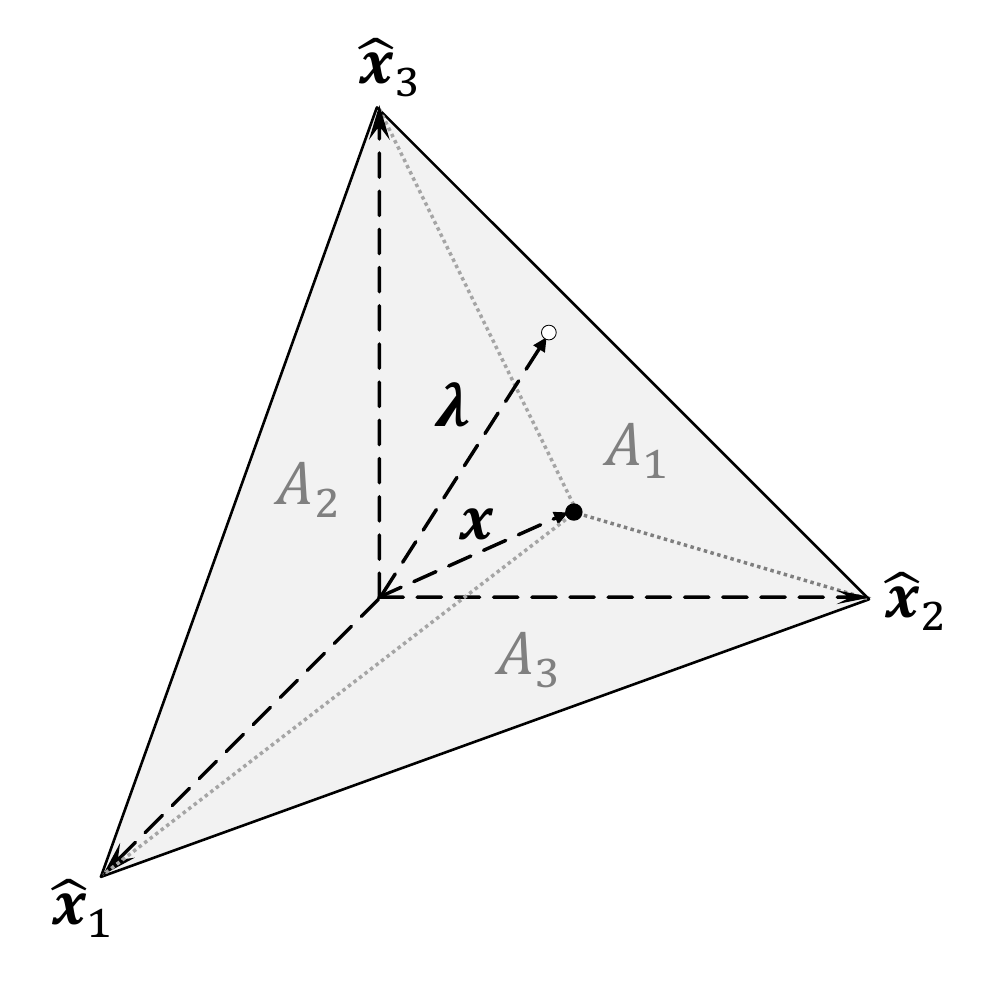}
\caption{A $2$-dimensional triangular membrane attached to the three vertex unit vectors $\hat{\bf{x}}_1$, $\hat{\bf{x}}_2$ and $\hat{\bf{x}}_3$, with the three disjoint convex regions $A_1$, $A_2$, and $A_3$, generated by the presence of the particle in $\bf{x}$ (the ``tension lines'' of demarcation between the three regions correspond to the clear  dashed lines in the drawing). The vector \mbox{\boldmath$\lambda$}, here in region $A_1$, indicates the point where the elastic membrane breaks.
\label{triangolo}}
\end{figure}

\begin{figure}[!ht]
\centering
\includegraphics[scale =.45]{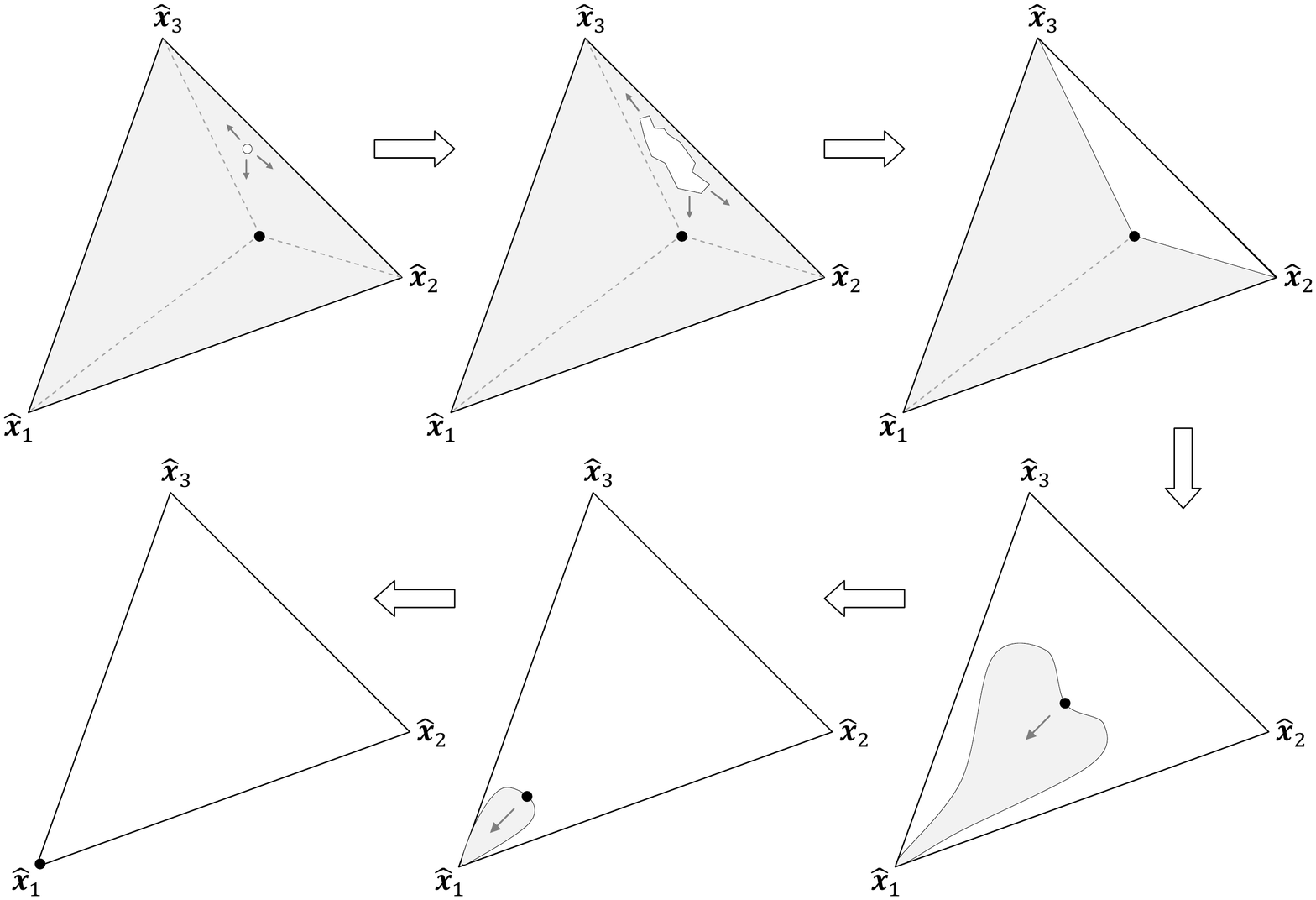}
\caption{The breaking of the elastic membrane (in grey color) in a $e_{\{1\}\{2\}\{3\}}$ measurement proceeds in two steps: first the membrane collapses, within the boundaries of the convex region containing the initial breaking point (here $A_1$), then, as soon as it loses the anchor points associated with this region, it shrinks towards the remaining anchor point, bringing with it the point particle (here to position $\hat{\bf{x}}_1$).  
\label{breakingprocess}}
\end{figure}

When this happens, three different disjoint convex regions $A_1$, $A_2$ and $A_3$ can be distinguished on the membrane's surface, delimited by the three ``tension lines'' which connect  $\bf{x}$ to the  vertex points of $S_2$ (see Fig.~\ref{triangolo}).
Then, after some time the elastic membrane breaks, at some unpredictable point \mbox{\boldmath$\lambda$} (see Fig.~\ref{triangolo}). If $\mbox{\boldmath$\lambda$}\in A_1$, then the tearing  propagates inside the entire region $A_1$, but not in the other two regions $A_2$ and $A_3$ (due to the presence of the tension lines), causing also its $2$ anchor points $\hat{\bf{x}}_2$ and $\hat{\bf{x}}_3$ to  tear away (from a physical point of view, the ``collapse'' of the membrane in region $A_1$ can be understood as a sort of explosive-like reaction of disintegration of its atomic constituents). Once the membrane is detached from the two above mentioned anchor points, being elastic, it  contracts toward the remaining anchor point $\hat{\bf{x}}_1$, drawing in this way the point particle, which is attached to it, to the same final position (see Fig.~\ref{breakingprocess}).
Similarly, if $\mbox{\boldmath$\lambda$}\in A_2$, the final state of the particle will be $\hat{\bf{x}}_2$, and if $\mbox{\boldmath$\lambda$}\in A_3$, the final state of the particle will be $\hat{\bf{x}}_3$. 

As for the previous description of the one-dimensional elastic band, we can observe that to each breaking point $\mbox{\boldmath$\lambda$}\in S_2$, corresponds a specific interaction between the particle and the elastic membrane, drawing the former  to its final state. In other terms, the measurement $e_{\{1\}\{2\}\{3\}}$ is formed by a collection of potential pure measurements, only one of which is each time actualized when the elastic breaks. Again, we observe that  all these pure measurements are deterministic, with the exception of those with a \mbox{\boldmath$\lambda$}  at the boundaries of two (or three) regions, as in this case it  remains indeterminate which region will actually disintegrate. But  of these special \mbox{\boldmath$\lambda$} we do not have to worry, as they are of zero measure in the determination of the transition probabilities. 

Following the same logic as for the two-outcome case, we have for the transitions  ${\bf{x}}\to \hat{\bf{x}}_{i}$, $i=1,2,3$, the probabilities $P({\bf{x}}\to \hat{\bf{x}}_{1}) = P(A_i) = {\mu_L(A_i)\over \mu_L(S_2)} = {2\over \sqrt{3}} \mu_L(A_i)$, were for the last equality we have used the fact that the area $\mu_L(S_2)$ of an equilateral triangle $S_2$ of side $\sqrt{2}$, is ${\sqrt{3}\over 2}$. To calculate the area of the triangle $\mu_L(A_i)$, we observe that its base is $\sqrt{2}$ and its height $h^i= \sqrt{3\over 2}x_i$, so that its area is ${\sqrt{2} h^i\over 2}={\sqrt{3}\over 2} x_i$. Thus, in accordance with (\ref{Born}), we obtain 
\begin{eqnarray}
\label{x1integral-z-u3}
P({\bf{x}}\to \hat{\bf{x}}_{i})= P(A_i) = {\mu_L(A_i)\over \mu_L(S_2)} = x_i, \quad i=1,2,3.
\end{eqnarray}
In other terms, the uniform membrane measurement $e_{\{1\}\{2\}\{3\}}$  is  isomorphic to the measurement of an non-degenerate observable $A$, in a three-dimensional complex Hilbert space ${\mathcal H}_3$, if we  represent the quantum state vector $|\psi\rangle = \sqrt{x_1}e^{i\alpha_1}|a_1\rangle +\sqrt{x_2}e^{i\alpha_2}|a_2\rangle +\sqrt{x_3}e^{i\alpha_3}|a_3\rangle \in {\mathcal H}_3$, by a vector ${\bf{x}} = x_1 \hat{\bf{x}}_1 + x_2 \hat{\bf{x}}_2 + x_3 \hat{\bf{x}}_3$, whose components are precisely the transition probabilities (see Sec.~\ref{Quantum Probabilities for a Single Observable}).
\\
\vspace{-0.3cm}
\\
\emph{The $N=3$ degenerate case, with two outcomes}
\\
As we previously mentioned, other typologies of measurements are possible with a two-dimensional membrane, that we have denoted $e_{\{1,2\}\{3\}}$, $e_{\{1,3\}\{2\}}$ and $e_{\{2,3\}\{1\}}$. Let us consider $e_{\{1,2\}\{3\}}$, the description of the other two measurements being similar. A measurement $e_{\{1,2\}\{3\}}$ corresponds to an experimental situation such that the experimenter decides not to discriminate between the two outcomes $\hat{\bf{x}}_1$ and $\hat{\bf{x}}_2$ (degenerate measurement). Therefore, the measurement only has two possible outcomes. To perform $e_{\{1,2\}\{3\}}$, the experimenter proceeds as follows. Once s/he has applied the uniform breakable membrane on $S_2$, s/he adds a highly reactive substance along the common boundary between $A_1$ and $A_2$. The effect of this special substance is twofold: (1) it produces the effective fusion of the two regions in a single region $A_{\{1,2\}}=A_1\cup A_2$, in the sense that if the membrane breaks in a point belonging, say, to $A_1$, the tearing  now propagates also across the boundary with $A_2$ (because of the presence of the reactive substance), causing the collapse of the entire region $A_{\{1,2\}}$; (2) it causes the detachment of the common anchor point $\hat{\bf{x}}_3$ before the other two anchor points $\hat{\bf{x}}_1$ and $\hat{\bf{x}}_2$.

\begin{figure}[!ht]
\centering
\includegraphics[scale =.45]{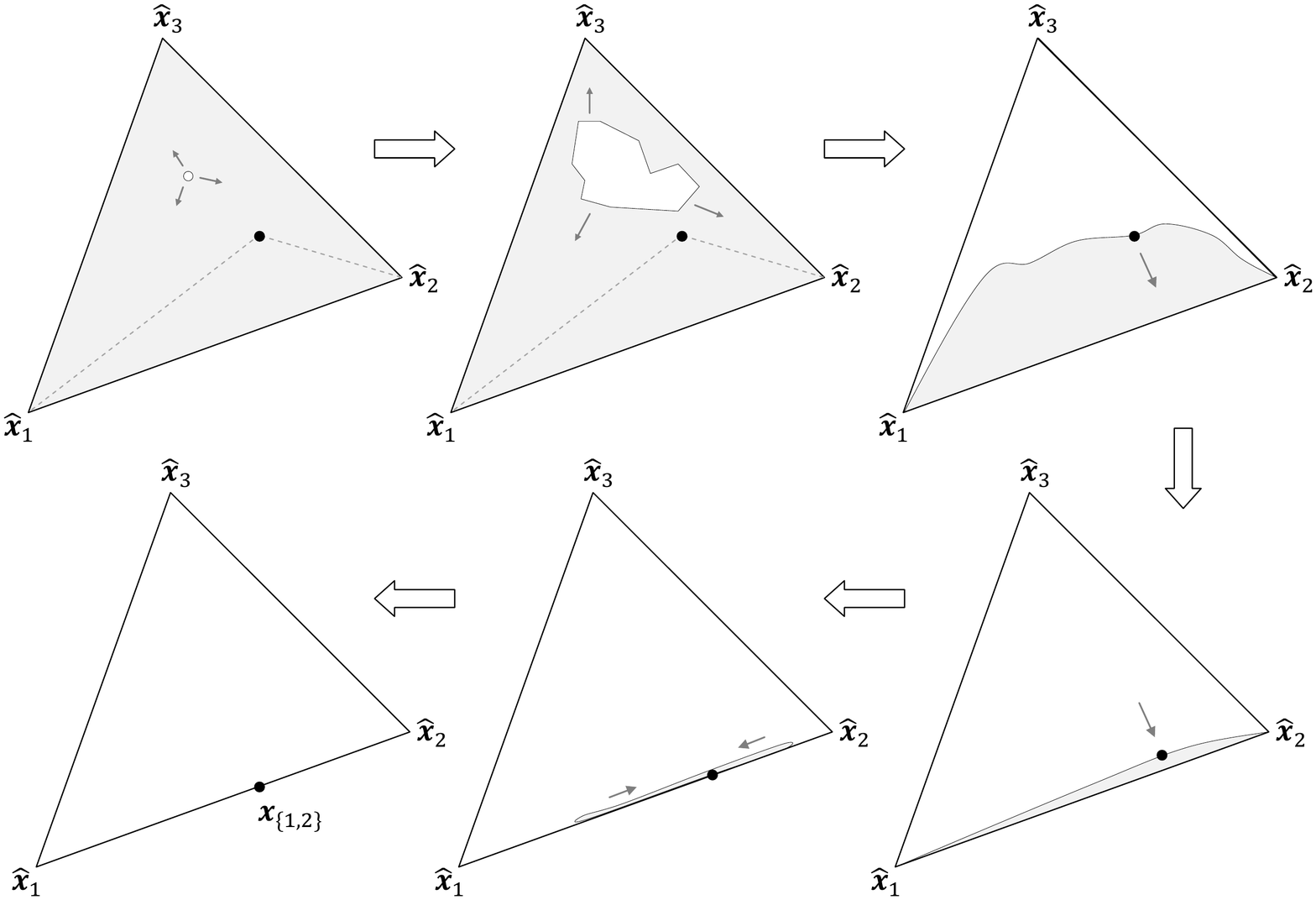}
\caption{The breaking of the elastic membrane (in grey color), when the two regions $A_1$ and $A_2$ are fused into a single region $A_{\{1,2\}}$, during a  $e_{\{1,2\}\{3\}}$ measurement. Here the process is represented in the case where the initial breaking point is in $A_2$. Firstly, the membrane in $A_{\{1,2\}}$ collapses, causing the common anchor point $\hat{\bf{x}}_3$ to detach and the particle to be drawn to position ${\bf x}_{\{1,2\}}$; then, also the two anchor points $\hat{\bf{x}}_1$ and $\hat{\bf{x}}_2$ detach, simultaneously, causing the membrane to shrink toward the particle, without affecting its acquired position.  
\label{breakingprocessdegenerate}}
\end{figure}

This means that, prior to the final detachment of the two anchor points  $\hat{\bf{x}}_1$ and $\hat{\bf{x}}_2$, because of the advanced detachment of anchor point $\hat{\bf{x}}_3$, the contraction of the elastic membrane will cause the particle to be drawn to point (see Fig.~\ref{breakingprocessdegenerate}):
\begin{eqnarray}
\label{vectorxmtriangle}
{\bf x}_{\{1,2\}} = {x_1 \over x_1 + x_2}\,\hat{\bf{x}}_1 +    {x_2 \over x_1 + x_2}\,\hat{\bf{x}}_2.
\end{eqnarray}
Then,  also the remaining two anchor points  $\hat{\bf{x}}_1$ and $\hat{\bf{x}}_2$  detach, and we assume they do so almost simultaneously, so that the membrane  contracts toward the particle, without affecting its acquired position ${\bf x}_{\{1,2\}}$, which therefore constitutes its final state, i.e., the outcome of the measurement. On the other hand, if the membrane breaks in $A_3$, then only that region collapses, producing the final outcome $\hat{\bf{x}}_3$ (as in the $e_{\{1\}\{2\}\{3\}}$ measurement). So, when performing $e_{\{1,2\}\{3\}}$, we have only two possible transitions: ${\bf{x}}\to {\bf x}_{\{1,2\}}$ and ${\bf{x}}\to \hat{\bf{x}}_{3}$, and the associated probabilities are:
\begin{eqnarray}
P({\bf{x}}\to {\bf x}_{\{1,2\}})=  {\mu_L(A_{\{1,2\}})\over \mu_L(S_2)}  =   x_1 + x_2, \quad P({\bf{x}}\to \hat{\bf{x}}_{3})= {\mu_L(A_3)\over \mu_L(S_2)}= x_3.
\end{eqnarray}

In other terms, the measurement $e_{\{1,2\}\{3\}}$  is  isomorphic to the measurement of a degenerate three-dimensional observable $A=a_{\{1,2\}} (P_1 +P_2) + a_3 P_3$, with $P_i=|a_i\rangle\langle a_i|$, $i=1,2,3$, where the possible post-measurement states are $|\psi_{\{1,2\}}\rangle = \sqrt{x_1 \over x_1 + x_2} e^{i\alpha_1} |a_1\rangle + \sqrt{x_2 \over x_1 + x_2}e^{i\alpha_2} |a_2\rangle$ and $|a_3\rangle$, and are represented by vectors ${\bf x}_{\{1,2\}}$ and $\hat{\bf{x}}_{3}$, in $S_2$, respectively. And similarly -- \emph{mutatis mutandis} --  for the measurements $e_{\{1,3\}\{2\}}$ and $e_{\{2,3\}\{1\}}$.
\\
\vspace{-0.3cm}
\\
\emph{The general $N$-outcome case}
\\
It is straightforward to generalize the working of the UTR-model to the case of an arbitrary number $N$ of outcomes.  The material point particle then lives in $\mathbb{R}^{n}$, with $n\geq N$,  
and to perform a (non-degenerate) measurement $e_{\{1\}\cdots\{N\}}$, a  uniform and breakable $(N-1)$-dimensional hypermembrane is stretched  over the hypersurface $S_{N-1}$ of a $(N-1)$-dimensional simplex generated by $N$ orthonormal vectors $\hat{\bf x}_1, \dots, \hat{\bf x}_{N}$, and attached to its $N$ vertex points. Once the  hypermembrane is in place, the particle, by moving deterministically towards it (along a trajectory that is not important here to specify, which will depend on the structure of the state space; see the discussion at the end of Sec.~\ref{nonhilbert}, for the case of a Hilbertian state space), sticks to it at a particular point: 
\begin{eqnarray}
\label{vector-N}
{\bf x}=\sum_{i\in I_{N}} x_i \hat{\bf x}_i, \quad  \sum_{i\in I_{N}} x_i = 1, \quad I_{N} \equiv \{1,\dots, N\},
\end{eqnarray}
which defines the state of the particle on the hypermembrane.

This gives rise to $N$ ``tension lines,'' connecting  ${\bf x}$ to the different vertex points $\hat{\bf x}_1, \dots, \hat{\bf x}_{N}$, defining in this way $N$ disjoint regions $A_i$, such that $S_N=\cup_{i\in I_{N}}A_i$ ($A_i$ is the convex closure of $\{\hat{\bf x}_1, \dots, \hat{\bf x}_{i-1}, {\bf x}, \hat{\bf x}_{i+1}, \dots, \hat{\bf x}_{N}\}$). Then, after some time the hypermembrane breaks, at some point ${\mbox{\boldmath$\lambda$}} =\sum_{i\in I_{N}} \lambda_i \hat{\bf x}_i$, $\sum_{i\in I_{N}} \lambda_i = 1$. If $\mbox{\boldmath$\lambda$}\in A_i$, for a given $i\in I_{N}$, then $A_i$ collapses, causing its $N-1$ anchor points $\hat{\bf x}_j$, $j\neq i$, to tear away. So, if  $\mbox{\boldmath$\lambda$}\in A_i$, the elastic hypermembrane  contracts toward point $\hat{\bf x}_{i}$, that is, toward the only point at which it remained attached, pulling in this way  the particle into that position. In other terms, the process produces the transition ${\bf x}\to \hat{\bf x}_{i}$, and the probability of such process is $P({\bf x}\to \hat{\bf x}_{i})= {\mu_L(A_i)\over \mu_L(S_{N-1})}$. Generalizing the previous reasoning for the three-outcome case (see Appendix~\ref{Uniform}), one can show that, for all $i\in I_N$:
\begin{eqnarray}
\label{transitionprobabilitiesnondeg}
P({\bf x}\to \hat{\bf{x}}_{i}) =  {\mu_L(A_i)\over \mu_L(S_{N-1})} = x_i, 
\end{eqnarray}
showing that  the measurement  $e_{\{1\}\cdots\{N\}}$  is  isomorphic to the measurement of an non-degenerate observable (\ref{observable-hilbert}), in a $N$-dimensional complex Hilbert space ${\mathcal H}_N$, if we  represent the quantum state vector (\ref{state-hilbert}) by the vector (\ref{vector-N}), whose components are precisely the transition probabilities (\ref{transitionprobabilitiesnondeg}).

We now also consider the more general class of measurements $e_{I_{M_1}\cdots I_{M_n}}$, $n=1,\dots,N$, corresponding to   situations where we have $n$ different subsets $I_{M_k}$ of $I_{N}$, $k=1,\dots,n$, $\sum_{k=1}^n M_k= N$, so that for each $k$, all regions $A_i$ having their indices in $I_{M_k}$ are fused together, and form a single structure $A_{I_{M_k}}=\cup_{i\in I_{M_k}}A_i$. In accordance with our previous description, the practical fusion of these regions is realized through the application of a special reactive substance at their common boundaries, so that the entire region $A_{I_{M_k}}$  collapses, whenever a breaking point manifests in one of its subregions. This produces first the disconnections of all anchor points shared by these subregions, causing the particle to be drawn by the elastic hypermembrane to position:
\begin{eqnarray}
\label{vectorxm}
{\bf x}_{I_{M_k}}=\sum_{i\in {I_{M_k}}} \left({x_i\over \sum_{i\in{I_{M_k}}x_i}}\right)\, \hat{\bf x}_{i},
\end{eqnarray}
and subsequently, because of the further simultaneous detachment of the remaining anchor points,  the entire hypermembrane  shrinks in the direction of the particle, without affecting its  acquired position ${\bf x}_{I_{M_k}}$. For a general $e_{I_{M_1}\cdots I_{M_n}}$-measurement, we thus have $n$ different possible outcomes ($1\leq n\leq N$), associated with the $n$ points ${\bf x}_{I_{M_k}}$, $k=1,\dots,n$, and  the transition probabilities are:
\begin{eqnarray}
\label{uniform-general}
P({\bf x}\to{\bf x}_{I_{M_k}})= {\mu_L(A_{I_{M_k}})\over \mu_L(S_{N-1})}  = \sum_{i\in I_{M_k}} x_i.
\end{eqnarray}
In view of (\ref{totalprob=1-degenerate}), a measurement  $e_{I_{M_1}\cdots I_{M_n}}$  is  therefore isomorphic to that associated with a degenerate observable (\ref{observable-hilbert-degenerate}), in a $N$-dimensional complex Hilbert space ${\mathcal H}_N$, where the possible post-measurement states are given by (\ref{new-state}), and are represented  in  ${\mathbb R}^N$ by the $n$ vectors (\ref{vectorxm}). For $n=1$, we have a single outcome, and the experiment is trivial, whereas for $n=N$ we recover the special case of the measurement $e_{\{1\}\cdots\{N\}}$, isomorphic to a non-degenerate observable (\ref{observable-hilbert}).
\\
\vspace{-0.3cm}
\\
\emph{A psychological mechanism}
\\
Before proceeding to the next section, where the UTR-model will be used to discuss the phenomenon of entanglement, it is important to spend a few more words on the membrane mechanism that we have described. The reader may indeed wonder what are the reasons behind our choice of using, as a specific realization of the measurement process, the dynamics of a breakable elastic membrane. Why should one be interested in such particular realization, and can one find a simpler representation for the probabilities? Also, is the model compatible with what we intuitively know about the functioning of a human cognitive process? 

In that respect, the important question to ask is the following: Is it possible to derive the quantum probabilities in terms of a consistent mechanism, able to describe general measurements, possibly degenerate, having an arbitrary number of outcomes? In this article, and in its second part \citep{AertsSassolideBianchi2014a}, we show that such a mechanical model can be constructed in terms of elastic breakable membranes (or better, hypermembranes), and this is per se already an interesting and unexpected result. However, is the dynamics of the membrane the only one one can use to represent the unfolding of a quantum (or quantum-like) measurement process? Do we have fundamental reasons for the utilization of membranes, instead of other systems? 

Here it is important to distinguish two different levels in the modelization. The first one is the choice of the geometric structure of the simplexes. This is really the fundamental part, as is clear that simplexes are the natural mathematical objects to be used to represent quantities, like probabilities, that sum to $1$. In other terms, all possible probabilities associated with an experiment with $N$ possible outcomes will naturally fill a $(N-1)$-dimensional simplex. The second level, less fundamental, is the description of a mechanism that can explain how a point on such simplex, representing the state of the system under investigation, can move from that position to one of its possible final states, which in the case of non-degenerate measurements are the $N$ vertices of the simplex. In that respect, our description of the dynamics of a breaking membrane can certainly be replaced, in principle, by some other descriptions. However, it is essential for the adopted mechanism to take into account the fact that the size (the Lebesgue measure) of the subregions formed by the on-membrane point particle, representative of the state, must be proportional to the probabilities of the different outcomes. The mechanism of the breaking membrane takes this fact into account in a very simple and natural way, and we haven't been able to imagine any simpler description. 

It should also be observed that the abstract breaking mechanism of the membrane is in fact very general and also a good metaphor of what we humans can intuitively feel when confronted with decisional contexts. In other terms, the membrane mechanism can certainly also be understood as a representation of an inner psychological mechanism. We can in fact consider that when a human subject is confronted with a question (and more generally with a decision), and an associated set of $N$ possible answers, this will automatically build a mental (neural) state of equilibrium, which results from the balancing of the different tensions between the initial state of the concept subjected to the question, and the available mutually excluding answers that compete with each other. The elastic membrane can then be seen as a convenient way to give shape to such a mental state of equilibrium, characterized by the presence of competing ``tension lines'' going from the specific position of the point particle on the membrane, representative of the initial state, to the $N$ vertices of the simplex, representative of the different possible answers. 

Still in accordance with what we can subjectively perceive, at some moment this mental equilibrium will be disturbed, in a non-predictable way, and the disturbance will cause an irreversible process during which, very quickly, the initial conceptual state will be drawn to one of the possible answers. This is represented in the model by the random breaking point on the membrane, which by collapsing also breaks the tensional equilibrium that had previously been built. This \emph{tension-reduction process}, however, will not always result in a full resolution of the conflict between all the competing answers. There are contexts such that the state of the system can be brought into another state of equilibrium, between a reduced set of possibilities. These sub-equilibriums are represented in our model by the different possible lower-dimensional sub-simplexes of the $(N-1)$-dimensional simplex, and describe those outcomes of degenerate measurements associated with degenerate eigenvalues.

Having said this, we conclude this section by remarking that the tension lines giving rise to the different convex regions of the membrane, are all formed by indeterministic hidden-measurement interactions, that is, by pure measurements giving rise to conditions of unstable equilibrium. It is interesting to note that it is the very existence of these indeterministic pure measurements that creates the different possibilities. On the other hand, the probability associated with these different possibilities, or outcomes, do not directly depend on these indeterministic pure measurements, but on the deterministic ones, which are contained inside the convex regions. This because, as already emphasized, the pure measurements associated with the ``tension lines'' are of zero measure, and cannot contribute to the value of the different probabilities. So, to put it in a different way, the tensions building the mental membrane's equilibrium are associated with unstable, indeterministic processes; these processes are at the origin of the different possibilities, but not of the values of the probabilities associated with them.

\vspace{-0.4cm}
\section{Entanglement in the UTR-model}
\label{Entanglementrhomodel}
\vspace{-0.4cm}

In this section, we exploit the UTR-model representation to gain some insight into the phenomenon of entanglement. In Section~\ref{Quantum Probabilities for a Single Observable} we have considered the example of a  compound system made of two entities, which can either be in a product (non-entangled) or non-product (entangled) state, and we have shown that the difference between non-entangled and entangled states is that for the former the probability for the outcome of a coincident product measurement  on the entity consisting of the compound of both entities, associated with an observable of the tensor product form $A_1\otimes A_2$, is equal to the product of the probabilities of these outcomes when the same two measurements are conducted separately on each entity, by means of the observables $A_1\otimes \mathbb{I}$ and  $\mathbb{I}\otimes A_2$, whereas for the latter this can never be the case. 

\begin{figure}[!ht]
\centering
\includegraphics[scale =.45]{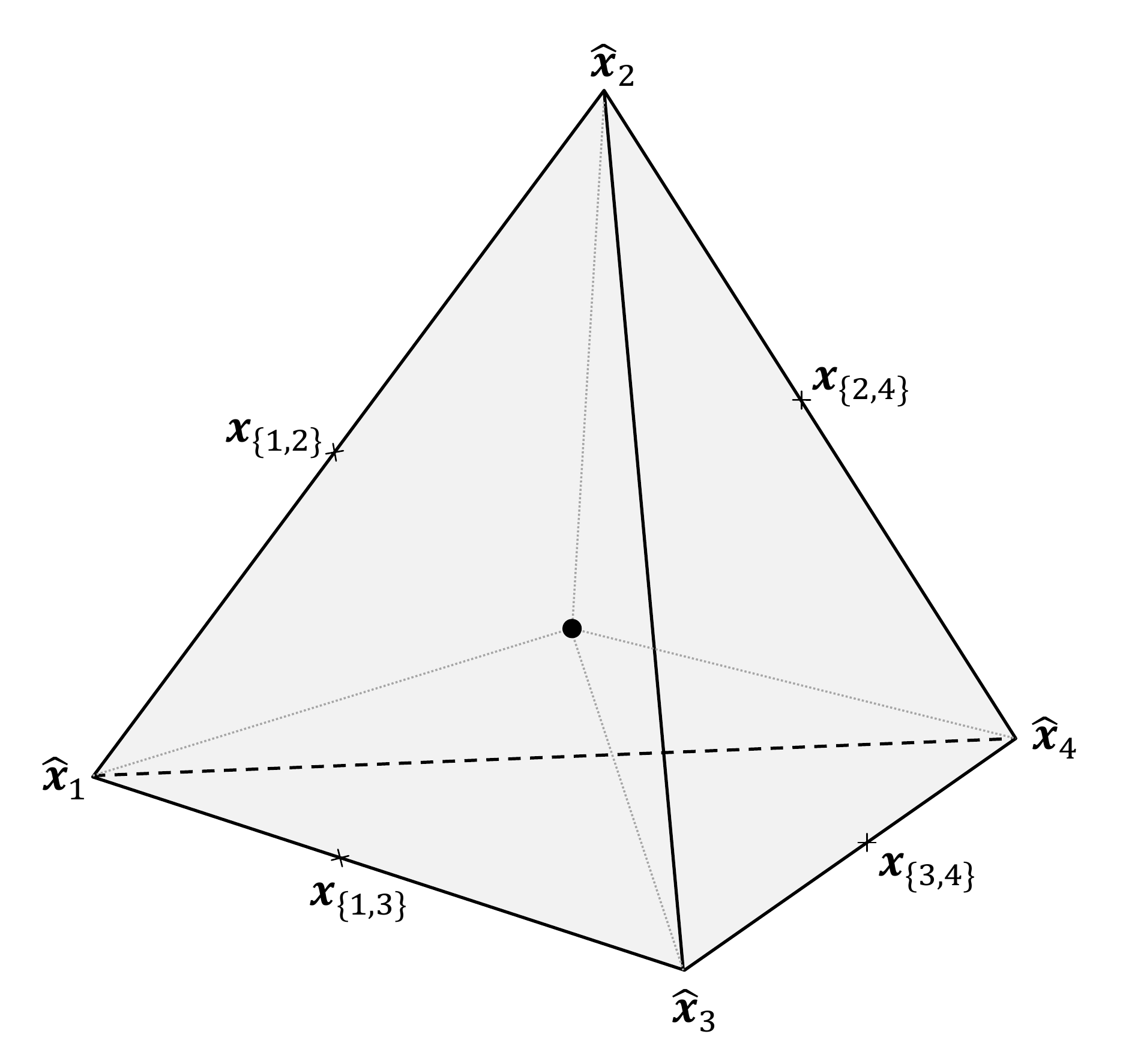}
\caption{A tetrahedron filled with an ``elastic gel'' (in grey color) describes in the UTR-model measurements with up to four different possible outcomes. One can see that the particle immersed in the volume of the tetrahedron defines four different convex regions (indicated with clear dashed lines), each one opposed to a different vertex of the volume, corresponding to the four outcomes of the $e_{\{1\}\{2\}\{3\}\{4\}}$ measurement. In the picture the two states ${\bf x}_{\{1,2\}}$ and ${\bf x}_{\{3,4\}}$, corresponding to the two possible outcomes of the (one-entity, degenerate) measurement $e_{\{1,2\}\{3,4\}}$, as well as the two  states ${\bf x}_{\{1,3\}}$ and ${\bf x}_{\{2,4\}}$, corresponding to the two possible outcomes of the (one-entity, degenerate) measurement $e_{\{1,3\}\{2,4\}}$, are also represented.
\label{thetraedron}}
\end{figure}

In the UTR-model, a four-outcome system can still be fully visualized, exploiting the fact that a three-dimensional hypermembrane $S_3$ can be represented in ${\mathbb R}^3$ as the volume of a tetrahedron (see Fig.~\ref{thetraedron}).
Following the notation of Section~\ref{Quantum Probabilities for a Single Observable}, we have that a measurement $A_1\otimes \mathbb{I}$  on the first entity corresponds in the UTR-model to  a measurement $e_{\{1,2\}\{3,4\}}$, and a measurement $\mathbb{I}\otimes A_2$  on the second  entity to a measurement $e_{\{1,3\}\{2,4\}}$. It is interesting to observe that when we perform these measurements one after the other, in whatever order, we obtain exactly the same result of the joint measurement $e_{\{1\}\{2\}\{3\}\{4\}}$, compatibly with the fact that  $A_1\otimes \mathbb{I}$ and $\mathbb{I}\otimes A_2$ commute. 

To see this, let us assume that we have performed first, say,  $e_{\{1,3\}\{2,4\}}$. The outcome ${\bf x}_{\{1,3\}}$ can then be obtained with probability $x_1+x_3$, whereas the outcome ${\bf x}_{\{2,4\}}$ can be obtained with probability $x_2+x_4$. Assuming for instance that we have obtained ${\bf x}_{\{1,3\}}$,  a further measurement  $e_{\{1,2\}\{3,4\}}$ will then produce either outcome $\hat{\bf x}_1$, with probability ${x_1\over x_1 +x_3}$, or outcome $\hat{\bf x}_3$, with probability ${x_3\over x_1 +x_3}$. Therefore, the probability that two sequential measurements   give outcome $\hat{\bf x}_1$ is given by the product  $(x_1+x_3)({x_1\over x_1 +x_3})=x_1$, and in the same way the probability that two sequential measurement  give outcome $\hat{\bf x}_3$ is $(x_1+x_3)({x_3\over x_1 +x_3})=x_3$. Reasoning in a similar way, if we assume that $e_{\{1,3\}\{2,4\}}$ has yielded instead ${\bf x}_{\{2,4\}}$, a further measurement  $e_{\{1,2\}\{3,4\}}$ will now produce either outcome $\hat{\bf x}_2$, with probability ${x_2\over x_2 +x_4}$, or outcome $\hat{\bf x}_4$, with probability ${x_4\over x_2 +x_4}$, so that the probabilities to obtain $\hat{\bf x}_2$ or $\hat{\bf x}_4$ in the  sequential measurement are  $(x_2+x_4)({x_2\over x_2 +x_4})=x_2$ and $(x_2+x_4)({x_4\over x_2 +x_4})=x_4$, respectively. And of course the same  holds true if we perform first $e_{\{1,2\}\{3,4\}}$, and then $e_{\{1,3\}\{2,4\}}$.

What we have just shown is that in the UTR-model, isomorphically to what happens in the quantum Hilbertian formalism, when we perform two different but compatible (i.e., commutable) ``coarse-grained'' measurements, one after the other, we obtain a ``finer grained'' measurement, where a greater number of outcomes can be distinguished. Now, since the two measurements $e_{\{1,2\}\{3,4\}}$ and  $e_{\{1,3\}\{2,4\}}$ only have two outcomes, they can also be  described, individually,  using a one-dimensional elastic structure, instead of a three-dimensional one. For instance, the measurement $e_{\{1,3\}\{2,4\}}$, performed on a particle in state ${\bf x}$ by means of a three-dimensional hypermembrane (represented in Fig.~\ref{thetraedron} as an elastic ``gel'' filling the volume of a tetrahedron), is  isomorphic to a measurement performed using a one-dimensional elastic band,  stretched over the two points ${\bf x}_{\{1,3\}}$ and ${\bf x}_{\{2,4\}}$,  as is clear that ${\bf x}$ can also be written as ${\bf x} = (x_1 + x_3) {\bf x}_{\{1,3\}} + (x_2 + x_4) {\bf x}_{\{2,4\}}$, and similarly, the measurement $e_{\{1,2\}\{3,4\}}$ is isomorphic to a measurement  using a one-dimensional elastic stretched over the two points ${\bf x}_{\{1,2\}}$ and ${\bf x}_{\{3,4\}}$,  as is clear that we can also write  ${\bf x} = (x_1 + x_2) {\bf x}_{\{1,2\}} + (x_3 + x_4) {\bf x}_{\{3,4\}}$. 

However, it is not possible to use two one-dimensional elastic band measurements, in sequence, to mimic the effects of a three-dimensional structure. Certainly, in the special case of an entity in a product state, one can always consider the two entities forming the compound system separately, each one represented in its own one-dimensional simplex, and perform separate measurements on each of them, then  combine the   probabilities for the different  outcomes to deduce those associated with a joint measurement. But this cannot be done if the two-entity system is in an entangled state, as only a genuinely three-dimensional structure will be able to account for all the experimental possibilities. In other terms, apart special (trivial) cases, it will not be possible to combine two one-dimensional elastic bands, say inside the structure of a tetrahedron, to reproduce the effects of the two measurements $e_{\{1,2\}\{3,4\}}$ and $e_{\{1,3\}\{2,4\}}$ performed in sequence, i.e., the effects of the ``fine-grained'' measurement $e_{\{1\}\{2\}\{3\}\{4\}}$. 

This means that higher dimensional structures can reproduce the behavior of lower dimensional ones, when (degenerate) sub-measurements are considered, but the converse is not true. This is an expression of what is called \emph{emergence}: when we combine two microscopic entities, like two electrons, in an entangled state, a genuine new entity emerges, which cannot be described in terms of the properties of the sub-entities forming the pair. Similarly, when two concepts are combined, a genuine new concept emerges, which cannot be understood only in terms of the two individual concepts of which it is the combination. 

Consider the following example, taken from \citet{AertsSozzo2011}, and further analyzed in \citet{AertsSozzo2012a}. The first entity is the concept \emph{Animal}, and a measurement of it consists in asking a subject to choose between the animal being a \emph{Horse} or a \emph{Bear}. This means that the concept \emph{Animal} is considered as a two-state system, and the above question is equivalent to an experiment performed with a one-dimensional elastic, with the two possible outcomes $\{H,B\}$. The second entity is the concept \emph{Acts}, and a measurement of it consists in asking a subject to choose between the act being either the emission of sounds like \emph{Growls}, or like \emph{Whinnies}. This means that the concept \emph{Acts} is again considered as a two-state system, and the question is  equivalent to an experiment performed with another one-dimensional elastic, with the two possible outcomes $\{G,W\}$. 

Consider then the compound system formed by both entities, in the state defined by their conceptual combination \emph{The Animal Acts}. This time we consider a joint measurement on both entities, which is about asking a subject to choose between the following four possibilities: \emph{The Horse Growls}, \emph{The Horse Whinnies}, \emph{The Bear Growls} and \emph{The Bear Whinnies}. This means that the two-concept system is  a four-state system, and that the above question is  equivalent to an experiment performed with a three-dimensional elastic hypermembrane (or a three-dimensional elastic ``gel'', in the tetrahedron representation), with the four possible outcomes $HG, HW, BG$ and $BW$. When data of the above three different measurements are collected \citep{AertsSozzo2011}, one finds that the probabilities do not obey relations (\ref{productrelations}), which means that the compound system formed by the two conceptual entities \emph{Animal} and \emph{Acts}, when in the state defined by the conceptual combination \emph{The Animal Acts}, is not in a product state, but in an entangled one. 

The entanglement of the two concepts is an expression of their connection through meaning. When a subject connects through meaning  \emph{Animal} and \emph{Acts}, s/he does so in a way that when, afterwards,  s/he considers exemplars of the combination \emph{The Animal Acts}, s/he will not refer back in a simple, combinatorial, ``logico-classical'' way to the exemplars of the individual concepts \emph{Animal} and \emph{Acts}. To express this in terms of the UTR-model analogy, s/he will not simply stretch a one-dimensional elastic over the \emph{Horse} and \emph{Bear} end points, and represent the state of \emph{Animal} as a point particle on it, and then do the same for the state of \emph{Acts}, which would correspond to another point particle on another one-dimensional elastic, stretched over the end points \emph{Growls} and \emph{Whinnies}. This type of ``parallel one-dimensional operations'' would be justified only if the two-concept system would be in a product state, corresponding to a situation where the concepts are combined without the creative emergent power of the human mind coming into play. Instead, what a human mind does, is to really build the equivalent of a three-dimensional elastic structure, and put the four different possible combinations $\{HG,HW,BG,BW\}$ as the four end-points of it (the four end points of a tetrahedron). By doing so, it attributes new weights to them, as ``good examples of'' \emph{The Animal Acts}. These new weights are certainly related, in some way, to the old weights (those associated with the individual concepts, i.e., which can be described by one-dimensional elastic structures), but cannot be derived from them in a simple combinatorial way. Indeed, all the experience of the subject, in her/his life, as regards to animals and the sounds they make,  comes into play in the determination of these new weights. This is a deep emergent creative process, expression of a dramatic change of the measurement context, whose increased level of potentiality needs additional dimensions to be described. A fact that is fully evidenced in the UTR-model, in the different possibilities offered by three-dimensional structures, in comparison to one-dimensional ones.

\vspace{-0.4cm}
\section{Representing the probabilities of a single measurement}
\label{RepresentingProbabilities}
\vspace{-0.4cm}

In the previous sections we have described the UTR-model by means of one of its possible mechanical realizations, which uses uniform hypermembranes that by breaking are able to draw a material point particle either to one of the $N$ vertices of a $(N-1)$-dimensional simplex $S_{N-1}$ (in case of a non-degenerate measurement), or to a point belonging to one of the lower dimensional sub-simplexes forming $S_{N-1}$ (in case of a degenerate measurement). We have described the  model mostly as a tool to represent quantum probabilities and understand how they emerge in a typical quantum measurement, showing that  a quantum measurement can be understood as an experiment involving a \emph{uniform} mixture of potential \emph{pure measurements}. These pure measurements are almost  classical, in the sense that, given the state of the point particle, almost all of them can be deterministically associated with a single outcome. However, since it is beyond the control power of the experimenter to know which specific pure measurement is each time actualized, outcomes can only be predicted in probabilistic terms. 

In other terms, the UTR-model is a model with a built-in mechanism able to explain the origin of quantum probabilities as the result of a uniform mixture of (hidden) pure measurements, which are available to be selected in a given experimental context, but in the ambit of a protocol which doesn't allow the experimenter to take any form of control over the selection mechanism  \citep{Aerts1986, Aerts1998, Aerts1999b, Coecke1995, SassolideBianchi2013a}. However, as we already mentioned, and as was noticed many years ago by one of us, this hidden-measurement approach is in fact an universal approach, in the sense that probabilities of whatever origin can always be explained and represented as being due to the presence of a lack of knowledge about the interaction between the experimental apparatus and the entity \citep{Aerts1994}. Of course, considering the correspondence that we have highlighted between  the probabilities described in the UTR-model, in terms of the uniform Lebesgue measure, and those of orthodox quantum mechanics, described by the Born rule, it is clear that also the Hilbert-model, with its scalar product, has to be considered a universal model for representing  arbitrary  probabilities appearing in nature, in a given measurement context. 

In other terms, the UTR-model, and equally so the Hilbert-model, are mathematical structures which can be used to represent in principle any probabilities emerging from the  interaction of two physical entities (the system under observation and the system which performs the observation). This is true, however, only if we consider a \emph{single} measurement situation. Indeed, if different measurements are considered, in a sequential process, the situation becomes much more complex and a more general framework is needed to describe the different probability models that can emerge from the interactions. This more general framework will be presented and analyzed in the next two sections of the article. 

For the moment, and considering our previous analysis, we can state the following representation theorem, valid for a single measurement situation
\citep{AertsSozzo2012a, AertsSozzo2012b}:\\

\vspace{-0.4cm}
\noindent {\bf Representation theorem} \emph{Given an arbitrary entity (e.g., a physical entity, or a conceptual entity) in a given state, and given a measurement, performed on it by means of another entity (e.g., a macroscopic measuring apparatus, or a human mind), with a set of possible outcomes $\{o_1,\dots, o_k\}$, with associated probabilities $\{p_1,\dots, p_k\}$, $p_1+\cdots +p_k=1$, $k\in {\mathbb N}$ (obtained as the limits of the relative frequency of the respective outcomes), then it is always possible to work out a representation of this experimental situation either in ${\mathbb R}^{N}$,  by means of the \emph{UTR-model}, with the probabilities given by the \emph{Lebesgue measure} of appropriately defined subregions of a $(N-1)$-simplex, or in a Hilbert space $\mathcal{H}_{N}\equiv {\mathbb C}^{N}$, with the probabilities given by the \emph{Born rule} of standard quantum theory, with $N$ an integer greater or equal to $k$}.
\\
\vspace{-0.4cm}

It is important to emphasize that although both the UTR-model and the Hilbert-model allow to universally represent a given single measurement situation, these two representations are certainly not equivalent. The advantage of the Hilbert-space representation is that it uses a manifest linear structure, which is particularly useful when one wants to see how probabilities associated with different states of the entity are related to each other, and describes these relations in terms of interference effects \citep{AertsSozzo2012b}. On the other hand, the advantage of the real-space representation of the UTR-model is that it doesn't assume linearity for the state space (a simplex is not a linear space), and therefore, from that point of view, it is a more general representation than the Hilbertian one. For instance, as we have seen in Section~\ref{Entanglementrhomodel}, the real-space representation of the UTR-model allows to identify and describe measurements on entangled states without the need of linearity  \citep{AertsSozzo2012a}. The UTR-model is a more general representation also because it allows for a finer description of the measurement process. Indeed, not only the state ${\bf x}$ of the entity prior to the measurement, and its final possible states\footnote{In our mechanical realization of the UTR-model, the outcomes of the measurements also correspond to states of the point particle. However, in a more abstract understanding of the model, it is not necessary to equate outcomes and states of the entity, in the sense that the model can be used also to represent situations where the state after the measurement cannot be necessarily identified.} $\hat{\bf x}_i$, are represented in the model, but also the states \mbox{\boldmath$\lambda$} of the measuring system, i.e., the pure measurements which are available to be actualized.

In the interpretation of the UTR-model that we have proposed in the previous sections, we have considered that the state of the system is given, i.e., that the system is in a pure state (specified by the vector ${\bf x}$), and that we are in the presence of a uniform mixture of pure measurements (specified by the equipotential \mbox{\boldmath$\lambda$} in the simplex $S_{N-1}$, i.e., by all the potential breaking points of the uniform hypermembrane). It is however interesting to observe that the model also allows for a symmetrical interpretation, in the sense that one can consider that only a single measurement interaction is available (a single \mbox{\boldmath$\lambda$}), whereas the state of the system would not be a priori given, but described by a uniform mixture. This is of course a very different dynamical picture. 

In our physical realization, it would correspond to the situation where the elastic hypermembrane, instead of being uniform, can only break in a single point \mbox{\boldmath$\lambda$}, and the experimenter, when applying the hypermembrane, has no possibility to know which will be the position of the point particle when sticking on it. For instance, one can assume that the particle would move so erratically that one could only predict such position in probabilistic terms, by means of a uniform probability density. In the ambit of a cognitive experiment with human subjects, we can think for instance of a situation where participants are asked to respond to a certain question in a deterministic way (according to some predetermined rule), but with the context of the question which is each time randomly changed, according to some uniform probability distribution. 

It is important here not to confuse a `mixture of states' with a `superposition of states,' as described in the linear Hilbert space model of quantum mechanics. Indeed, it is customary to affirm that, immediately before asking an experimental question, a quantum entity can be in an indefinite state, described by a superposition. This superposition, however, describes in the quantum formalism a superposition of \emph{final} possible states of the entity (the outcomes), and cannot in general be interpreted as a statistical mixture. In what we are considering here, the mixture of states  refers to a mixture of \emph{initial} states, in the sense that the experimenter, when performing the measurement, is not able to control, and therefore to know, which is the initial state of the entity. Therefore, even though s/he knows that the only available pure measurement produces (almost all the times) a final state in a perfectly deterministic way, s/he is unable to predict such final state, as s/he lacks knowledge about the initial condition of the measurement process. 

One can also interpret this situation in a reversed way, by considering that it is not the elastic hypermembrane which measures the state of the point particle, but the (now erratically moving) point particle which measures the ``breakability'' state of the hypermembrane. From that inverted viewpoint, the final positions of the point particle are to be interpreted as the final states of the hypermembrane. The mathematical description of this symmetric, complementary view of the UTR-model can be easily deduced by answering the following question: What are the points of $S_{N-1}$ representing the possible states of the particle that are all dragged to a same final state (outcome), when only a single pure measurement \mbox{\boldmath$\lambda$} is available, i.e., when the hypermembrane can only break in a single point \mbox{\boldmath$\lambda$}? 

\begin{figure}[!ht]
\centering
\includegraphics[scale =0.6]{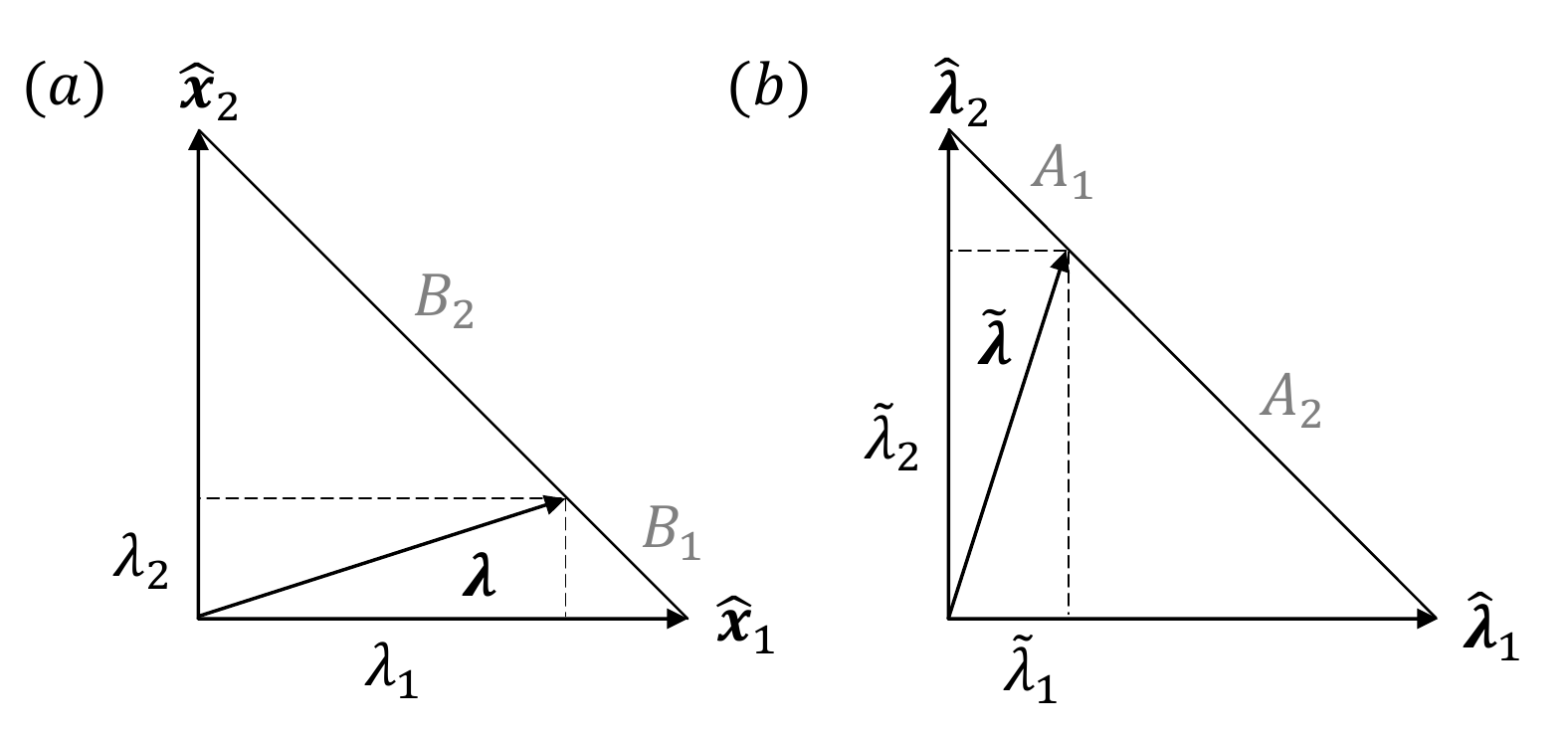}
\caption{A pure measurement represented in figure $(a)$ as a point \mbox{\boldmath$\lambda$}  in the 1-dimensional simplex generated by the final states of the measured entity, with the region $B_1$ (resp. $B_2$) corresponding to those initial states that are all changed by the interaction into the final state $\hat{\bf x}_1$ (resp. $\hat{\bf x}_2$). In figure $(b)$ the same pure measurement is represented as a point \mbox{\boldmath${\tilde \lambda}$} in the 1-dimensional simplex generated by the final states of the measuring entity, with the region $A_1$ (resp. $A_2$) corresponding to those initial states of the point particle that change the state of the measuring entity into the final state $\hat{\mbox{\boldmath$\lambda$}}_1$ (resp. $\hat{\mbox{\boldmath$\lambda$}}_2$).
\label{1-dimensionsbis}}
\end{figure}

For $N=2$, it is easy to see in Fig.~\ref{1-dimensionsbis} $(a)$ that all  states ${\bf x}$ of the particle belonging to  region $B_2$, bounded by vectors $\hat{\bf{x}}_2$ and \mbox{\boldmath$\lambda$}, will be drawn to $\hat{\bf x}_2$, whereas all states belonging to region $B_1$, bounded by vectors \mbox{\boldmath$\lambda$} and $\hat{\bf{x}}_1$, will be drawn to $\hat{\bf x}_1$. Therefore, the probability for a point particle whose initial position is uniformly randomly chosen, to be drawn to $\hat{\bf{x}}_1$, is $P(\to\hat{\bf{x}}_1)={\mu_L(B_1)\over \mu_L(S_1)}=\lambda_2$, whereas the probability to be drawn to $\hat{\bf{x}}_2$, is $P(\to\hat{\bf{x}}_2)={\mu_L(B_2)\over \mu_L(S_1)}=\lambda_1$. To make fully manifest the connection with the Born rule of quantum mechanics, also in this complementary view, one has to work directly in the ``state of the apparatus'' representation, instead of the ``state of the entity'' representation, by setting $\tilde{\lambda}_1=\lambda_2$ and $\tilde{\lambda}_2=\lambda_1$, so that the probabilities become (see Fig.~\ref{1-dimensionsbis} $(b)$): $P(\mbox{\boldmath${\tilde \lambda}$} \to\hat{\mbox{\boldmath$\lambda$}}_i)={\mu_L(A_i)\over \mu_L(S_1)}=\tilde{\lambda}_i$, $i=1,2$. Then, the Hilbert-model representation of these same probabilities can be obtained by means of the Born rule if one considers a state vector $|\phi\rangle = \sqrt{\tilde{\lambda}_1}e^{i\beta_1}|b_1\rangle +\sqrt{\tilde{\lambda}_2}e^{i\beta_2}|b_2\rangle \in {\mathcal H}_2$, which now describes the state of the measuring entity, instead of the state of the measured one. 

The above alternative scheme, which can be generalized to an arbitrary number $N$ of outcomes (see Appendix~\ref{Mixed}, for the $N=3$ case), highlights an interesting symmetry between the pure states of the measuring entity and the pure states of the measured entity. This symmetry tells us that probabilities associated with a given, single measurement situation, can be described by either assuming that the state of the  system is perfectly known, and the potentials acting on the system are the result of a uniform mixture of pure measurements, or by assuming that the state of the measuring system is perfectly known, and the potentials are the result of an unknown mixture of pure states of the measured system, and that the quantum mechanical Born rule is compatible, from a mathematical viewpoint, with both situations (in the ambit of a single measurement situation). 

In the next section we will explore more deeply the nature of the potentiality which emerges from the contact  between the observer and the observed, but to do so we first need to introduce an even more general model than the UTR-model, or the Hilbert-model. This will allow us to introduce the important notion of \emph{universal measurement}.

\vspace{-0.4cm}
\section{Averaging over non-uniform fluctuations}
\label{Non-uniform}
\vspace{-0.4cm}

In the previous sections we have described the UTR-model and have shown that, likewise  the Hilbert-model,  it can be understood as a ``universal probabilistic machine,'' in the sense that it corresponds to a mathematical structure able to represent every set of probabilities appearing in a given single measurement context, whatever its nature. More specifically, what we have done is to use a specific physical realization of the UTR-model to highlight the structure of the hidden dynamics which is inherent in a quantum measurement, explaining the emergence of probabilities as due to the presence of a ``region'' of potentiality between the measuring and measured systems, which can either originate from a lack of knowledge about the initial condition of the measured system, or as a lack of knowledge about the state of the measuring apparatus (i.e., about the pure measurement which is each time actualized). 

We want now to consider a more general class of measurements. To do so, we observe that in our discussion of the UTR-model, we have only considered \emph{uniform} elastic hypermembranes, and this is the reason why the different probabilities in the model were obtained by means of the Lebesgue measure. In other terms, so far we have implicitly considered that each hypermembrane is characterized by a uniform probability density $\rho_u$, describing how the elastic can break. But of course, we can imagine hypermembranes that can break in a variety of different ways, depending on how they have been manufactured and on the nature of the environment in which they are immersed. This amounts assuming that each elastic structure is characterized by a more general, not necessarily uniform, probability density $\rho : S_{N-1}  \to [0,\infty [$, describing the probabilities for the hypermembrane of breaking in the different regions of $ S_{N-1}$. Accordingly, the probability $P(A|\rho)$ for a $\rho$-hypermembrane (i.e., an hypermembrane characterized by the probability density $\rho$) to break in a given region $A$ of $S_{N-1}$,  is now given by the integral  $P(A|\rho)=\int_{A} \rho({\bf{y}})d{\bf{y}}$, which corresponds to the Lebesgue measure of $A$ only in the case $\rho$ would be uniform. Therefore, the different transition probabilities are now conditional to the  specific choice of the $\rho$-hypermembrane used to perform the measurement, i.e., 
\begin{eqnarray}
\label{rhoprobabilitymeasurement}
P({\bf x}\to \hat{\bf{x}}_{i}|\rho) =  P(A_i|\rho)=\int_{A_i} \rho({\bf{y}})d{\bf{y}}, \quad i\in I_N.
\end{eqnarray}
 
Clearly, similarly to the the UTR-model,  this more general description of a measurement, which we shall call the \emph{general tension-reduction} (GTR) model,  also allows for a full visualization of what goes on during the measurement process, in terms of the collapse of breakable structures, and the same  discussion presented in Sections~\ref{Thepolytopemodel} and~\ref{Entanglementrhomodel} can be repeated for non-uniform hypermembranes (i.e., non-uniform $\rho$). In other terms, uniform and non-uniform hypermembranes exemplify the same measurement paradigm, which is that of the \emph{hidden-measurement approach} \citep{Aerts1986, Aerts1998, Aerts1999b, SassolideBianchi2013a}, where the emergence of quantum and quantum-like processes is explained as the consequence of the presence of fluctuations in the measurement context. But, as we said, the interest of the GTR-model, in comparison to the (uniform) UTR-model, is in its ability of providing a much more general theoretical framework, able to describe different typologies of measurements, characterized by different forms of fluctuations, which can give rise to different probability models (as we will demonstrate in a very explicit way in the following section, in the two-outcome situation).

For instance, in the two-outcome case, the GTR-model allows for the description of elastic bands which can  uniformly break only in their central segment (the so-called $\epsilon$-model), a situation which can give rise to non-Kolmogorovian and non-Hilbertian probability models \citep{Aerts1986, Aerts1995, AertsSassolideBianchi2014a}, and if we let the length of such segment tends to zero, we find the special case of a probability density which is a Dirac delta-distribution $\rho(z) = \delta(z)$, describing a situation where only a single deterministic interaction (what we have called a pure measurement) would be available to be selected. This means that one can know in advance where exactly  the elastic will break, with certainty, and outcomes can be described by probabilities which can only take the values $1$ or $0$ (if we exclude the special case of a particle situated exactly on the breaking point), depending only on the initial state of the system.  

In a cognitive experiment, this would correspond to a situation where all the subjects would give the same fully predictable answers to the questions addressed to them, for example because they would have decided in advance to respond to them on the basis of a predetermined script. Since no genuine potentiality is involved in (almost) deterministic measurements of this kind, we can say that they  maximize the \emph{discovery} aspect, as they can only reveal what was already present (actual) before the execution of the measurement.\footnote{During a deterministic measurement the system will undergo, in general, a transition from an initial to a final state. Therefore, we can certainly affirm that the measurement creates a new state. However, since the process is fully predictable, we cannot consider this process as a process of \emph{actualization of a potential property}. It is only when some level of potentiality is actualized during an experiment, in a genuinely unpredictable way, that we can speak of the creation of a new property, and therefore affirm that there is an element of creation involved in the measurement (Sassoli de Bianchi, 2012).} 

Somehow opposite to (almost) deterministic pure measurements, maximizing the discovery aspect, the GTR-model is also able to describe what we may call ``solipsistic measurements,'' maximizing the creation aspect. As a simple example, consider a measurement carried out  with an elastic whose breakability is described by a double-Dirac distribution: $\rho(z) = a\delta(z-{1\over \sqrt{2}}) + b\delta(z+{1\over \sqrt{2}})$, $a+b=1$, $a,b>0$. This describes a situation where the elastic  can only break in its two end points, so that the probabilities of the two outcomes will not depend anymore on the  specific state of the point particle. In other terms, we are here in a situation where the measurement reveals nothing about the state of the entity before the experiment, as is clear that only the structure of the hidden interactions is important to determine the value of the probabilities (hence the term ``solipsistic'' used to denote these experiments, to emphasize that they only tell us about the state of the observer, i.e., of the measuring apparatus, and not about the state of the observed system). 

Therefore, we can say that, opposite to deterministic measurements,  solipsistic measurements minimize the \emph{discovery} aspect and maximize the \emph{creation} aspect. In a cognitive experiment, this could correspond  to a situation where the subjects are totally insensitive to the way the questions are formulated, and respond to them in a genuinely unpredictable way, according only to the fluctuations of their state of mind. In other terms, in a pure solipsistic measurement the subjects would not be affected by the context defining the state of the concepts which are contained in the questions, so that their answers cannot reveal anything about the nature of the questions, but only about the nature of their state of mind at that moment. In that sense, we could say that solipsistic measurements can possibly modelize behaviors of subjects who, for whatever reasons, would understand language only at a very basic level (involving vocabulary and spelling, thus allowing them to respond to the  questions), but with difficulties in grasping more complex language structures, such as the symbolic and figurative contents.   

In between these two extremes, describing on one side measurements maximizing the discovery aspect, and on the other side measurements maximizing the creation aspect, all possible intermediary situations, mixing these two aspects in infinitely many different combinations, are possible and describable within the GTR-model. This because there are no specific restrictions in the choices of  the probability density $\rho$, which only needs to be an integrable function, and therefore can also be discontinuous, and include the limit case of distributions. In other terms, the model is very general, as it includes pure classic measurements and pure solipsistic measurements, with the pure quantum measurements somewhere in between, as well as all possible quantum-like measurements described by probability models which are different from the classical, quantum and solipsistic ones, thus corresponding to truly intermediary situations. 

Now, considering the existence of all these  different possible regimes of creation and discovery, corresponding to different degrees of availability of the hidden deterministic interactions that can be selected during a measurement and produce a specific outcome, it is natural to consider a more general class of measurements, which we denote \emph{universal measurements}, such that not only a hidden interaction would be actualized during their execution, i.e., a given breaking point of the elastic structure, but  an entire probability law $\rho$, from which a given breaking point would then be obtained. 

To put it differently, a universal measurement $e^{\rm{univ}}\equiv \{e^\rho\}$  is a meta-measurement where, at each measure, an entire measurement $e^\rho$, characterized by a probability density $\rho$, is actualized (in a randomly uniform way) and carried out. To put it even differently, a universal measurement describes a situation of very deep lack of knowledge regarding the experiment which is each time carried out, as it doesn't only describe a situation of lack of knowledge regarding the deterministic interaction which is actualized to produce the outcome, but also a situation of lack of knowledge regarding the very typology of experiment which is conducted, i.e., about the $\rho$ that characterizes it. As we discussed in the introduction, this is the kind of situation which risks to be the typical one in cognitive experiments performed with different subjects, as their different ``ways of choosing'' are  usually not discriminated, but averaged out in the final statistics of outcomes.

But how can we define a uniform randomization over the different possible probability densities? A possibility would be that of introducing a parameterization of the probability density $\rho$, by means of a finite set of parameters $\epsilon_i$, $i=1,\dots,n$,  then calculating the averages of the outcome probabilities over these parameters (performing a multiple integral over them), which would then correspond to the probabilities associated with a hypothetical universal measurement. The problem with this kind of strategy is that there is no natural procedure to introduce such parameterization. For instance, in the two-outcome case, one can certainly start by introducing a single parameter $\epsilon\in [0,1]$, and the parametrization: $\rho_\epsilon(z)={1\over \sqrt{2}\epsilon}\chi_{[-{\epsilon\over \sqrt{2}},{\epsilon\over \sqrt{2}}]}(z)$, where $\chi_{[-{\epsilon\over \sqrt{2}},{\epsilon\over \sqrt{2}}]}(z)$ is the characteristic function of the interval $[-{\epsilon\over \sqrt{2}},{\epsilon\over \sqrt{2}}]$, describing the region where the elastic is uniformly breakable (this is the parametrization chosen in the $\epsilon$-model). It is then not difficult to integrate over $\epsilon$ the probabilities (\ref{rhoprobabilitymeasurement}), and  calculate in this way average probabilities, but then why limiting the averaging only to symmetric uniformly breakable regions? Why not considering also asymmetric ones? 

For instance, instead of a single parameter, we could introduce two parameters $\epsilon_1, \epsilon_2$, with $\epsilon_1\in  [0,1]$ and $\epsilon_2\in  [-1+\epsilon_1,1-\epsilon_1]$, thus admitting more general elastics, unbreakable in their left part, from $-{1\over\sqrt{2}}$ to ${\epsilon_2-\epsilon_1\over\sqrt{2}}$, uniformly breakable from ${\epsilon_2-\epsilon_1\over\sqrt{2}}$ to ${\epsilon_2+\epsilon_1\over\sqrt{2}}$, and again unbreakable in their right part, from ${\epsilon_2+\epsilon_1\over\sqrt{2}}$ to ${1\over\sqrt{2}}$, described by the probability densities: 
$\rho_{\epsilon_1,\epsilon_2}(z)={1\over \sqrt{2}\epsilon_1}\chi_{[{\epsilon_2-\epsilon_1\over\sqrt{2}},{\epsilon_2+\epsilon_1\over\sqrt{2}}]}(z)$. But then, even in this very simple case of a probability density defined in terms of only two parameters, one immediately face a serious problem: Bertrand's paradox \citep{Bertrand1889}. Indeed, as emphasized more than a century ago by the French mathematician Joseph Bertrand, when the sample space of events is infinite, there is apparently no unambiguous way to define the term `at random'. For instance, considering the probabilities associated with the above defined $\rho_{\epsilon_1,\epsilon_2}$, how do we have to average them, in order to describe the situation of a random choice of an elastic characterized by a couple of parameters $(\epsilon_1,\epsilon_2)$? Just to give an example, we could decide to choose the couple $(\epsilon_1,\epsilon_2)$ at random in the triangle defined by the lines $\epsilon_1=-\epsilon_2 +1$, $\epsilon_1=\epsilon_2 +1$, and $\epsilon_1=0$. If we do so, we will find certain specific values for the average outcome probabilities, but one can invent many other ways of choosing $(\epsilon_1,\epsilon_2)$ at random, thus defining different uniform averages for the probabilities, and therefore obtain different numerical values for them. 

So, we apparently face here a double problem. $(1)$ The first one is that we don't seem to have any a priori criterion for deciding, on a physical or logical basis, how many continuous variables $\epsilon_i$ we should use to parameterize the probability density $\rho$, and how to do it. For instance the above $(\epsilon_1,\epsilon_2)$-model (which generalizes the one-parameter $\epsilon$-model), does not allow to represent the possibility of the previously mentioned solipsistic measurements, which therefore would be left out from the average. $(2)$ The second problem, specifically related to Bertrand paradox, is that even if we can describe a sufficiently general model, by means of a finite number of continuous parameters $\epsilon_i$, $i=1,\dots,n$, which would hypothetically be able to describe all relevant measurement situations, there would still be the problem of finding a non-ambiguous way of choosing at-random these parameters, in order to obtain a uniform average, as infinitely many uniform averages can a priori be defined, yielding different numerical values for the probabilities. 

This problem, which seems to prevent us from giving an unambiguous definition, and therefore attribute a clear meaning, to the notion of universal measurements, was already noticed by one of us in \citet{Aerts1998} and \citet{Aerts1999b}. There, it was observed that, despite this difficulty, there was nevertheless the possibility that a notion of universal measurement could be defined, with a clear physical and mathematical meaning. This possibility was indirectly suggested by a famous theorem of quantum mechanics: \emph{Gleason's theorem}. Indeed, being a universal measurement a measurement consisting of a huge, uniform average over all possible parameters distinguishing between different possible probability densities, it is certainly one of its remarkable properties that of being characterized by probabilities which, by definition, would only depend on the initial, preparation state, and the final, outcome state. But this is exactly the property of probabilities defined by means of the Born rule! More precisely, quoting from \citet{Aerts1998}:
\vspace{-0.25cm}
\begin{quote}
``Gleason's theorem states that `if the transition probability depends only on the state before the measurement and on the eigenstate of the measurement that is actualized after the measurement, then this transition probability is equal to the quantum transition probability'. But this Gleason property (dependence of the transition probability only on the state before the measurement and the eigenstate that is actualized after the measurement) is exactly a property that is satisfied by what we have called the `universal' measurements. Indeed, the transition probability of a universal measurement, by definition of this measurement, only depends on the state before the measurement and the actualized state after the measurement. Hence Gleason's theorem shows that the transition probabilities connected with universal measurements are quantum mechanical transition probabilities. We go a step further and want to interpret now the quantum measurements as if they are universal measurements. This means that quantum mechanics is the theory that describes the probabilistics of possible outcomes for measurements that are mixtures of all imaginable types of measurements. Quantum mechanics is then the first-order non classical theory. It describes the statistics that goes along with an at-random choice between any arbitrary type of manipulation that changes the state $p_v$ of the system under study into the state $p_u$, in such a way that we don't know anything of the mechanism of this change of state. The only information we have is that `possibly the state before the measurement, namely $p_v$, is changed into a state after the measurement, namely $p_u$'. If this is a correct explanation for quantum statistics, it explains its success in so many regions of reality, also concerning its numerical statistical predictions.''
\end{quote}
\vspace{-0.25cm}

What was only conjectured in \citet{Aerts1998}, we are now in a position to prove, thanks to a physically transparent and mathematically precise definition of what a universal measurement is. The formal proof of this result will not be given in the present article, but in its second part \citep{AertsSassolideBianchi2014a}, as the main scope of the present work is to explain the functioning of the UTR-model and introduce its non trivial GTR-model generalization, so as to allow for a correct contextualization of the result of the equivalence between the universal measurements and the measurements described by the (uniform) Lebesgue measure, which in turn are isomorphic  to the measurements described by the Born rule (when the states describing the entity under consideration admit a Hilbert space representation). 

For this, we need now to give a sufficiently general and consistent  definition of a universal measurement, which has to include in its average all possible measurements, but at the same time remain well posed, in the sense that it must not suffer from the ambiguities of typical Bertrand paradox situations, where the randomization process is not uniquely defined. In other terms, we need to find a general probability measure on the non-denumerable set of $N$-dimensional integrable generalized functions $\rho$, without being confronted with technical problems related to the foundations of mathematics and probability theory. This can be done by using the following strategy (for a demonstration of the following statements, we refer the reader to \citet{AertsSassolideBianchi2014a}): 

\vspace{0.1cm}
\noindent (1) First, one shows that any  probability density $\rho$ can be described as the limit of a suitably chosen sequence \emph{cellular probability densities} $\rho_{n_c}$, as the number of cells $n_c$ tends to infinity, in the sense that for every initial state ${\bf x}$ and final state ${\bf x'}$, we can always find a sequence of cellular $\rho_{n_c}$, such that the transition probability $P({\bf x}\to {\bf x'}|\rho_{n_c})$ tends to $P({\bf x}\to {\bf x'}|\rho)$, as $n_c\to\infty$. By a cellular probability density we mean a probability density describing a structure made of a total number $n_c$ of regular cells (of whatever shape), which  tessellate the hypersurface of the simplex $S_{N-1}$. These $n_c$ cells can only be of two sorts: uniformly breakable, or uniformly unbreakable.

\vspace{0.1cm}
\noindent (2) Thanks to the fact that a cellular probability density $\rho_{n_c}$ is only made of a finite number $n_c$ of cells, which can either be of the breakable or unbreakable kind, if we exclude the totally unbreakable case of a $\rho_{n_c}$ describing a structure only made of unbreakable cells (which would produce no outcomes in a measurement), we have that the total number of possible cellular $\rho_{n_c}$ is $C_{n_c}^0 + C_{n_c}^1 + C_{n_c}^2 + \cdots + C_{n_c}^{n_c} -1 = 2^{n_c}-1$. Therefore, for each $n_c$, we can unambiguously define the average probability: 
\begin{eqnarray}
\label{average1}
P({\bf x}\to {\bf x'}|n_c) \equiv {1\over 2^{n_c}-1}  \sum_{\rho_{n_c}}P({\bf x}\to {\bf x'}|\rho_{n_c}),
\end{eqnarray}
where the sum runs over all the possible $2^{n_c}-1$ cellular probability densities made of $n_c$ cells. Clearly, $P({\bf x}\to {\bf x'}|n_c)$ is the probability of transition ${\bf x}\to {\bf x'}$, when a cellular probability density $\rho_{n_c}$ (a cellular hypermembrane) is chosen uniformly at random. Since the total number of different possible $\rho_{n_c}$, for a given $n_c$, is finite, there are no ``Bertrand paradox'' ambiguities in the definition of the uniform average (\ref{average1}), which is therefore unique. 

Thanks to the above, we can now define a general universal measurement  $e^{\rm{univ}}$ (which can be either non-degenerate or degenerate, depending on whether the uniform average involves non-degenerate or degenerate measurements, distinguishing or not distinguishing all the possible $N$ outcomes) as follows:\\

\vspace{-0.4cm}
\noindent {\bf Definition (Universal Measurement)}. \emph{A measurement is said to be a \emph{universal measurement $e^{\rm{univ}}$} if the probabilities associated with all its possible transitions ${\bf x}\to {\bf x'}$ are the result of a uniform average over all possible measurements $e^\rho$, described by all possible probability densities $\rho$, as defined by the infinite-cell limit: 
\begin{eqnarray}
\label{average-limit}
P^{\rm{univ}}({\bf x}\to {\bf x'})=\lim_{n_c\to\infty} P({\bf x}\to {\bf x'}|n_c),
\end{eqnarray}
where $P({\bf x}\to {\bf x'})|n_c)$ is the average \emph{(\ref{average1})}. 
}
\\
\vspace{-0.4cm}

\noindent Thanks to the above definition, we are now in a position to enunciate the following theorem, which connects the universal measurements with the measurements described by the uniform Lebesgue measure \citep{AertsSassolideBianchi2014a}:\\

\vspace{-0.4cm}
\noindent {\bf Theorem (Universal $\Leftrightarrow$ Uniform)}. \emph{A universal measurement $e^{\rm{univ}}$ is probabilistically equivalent to a measurement $e^{\rho_u}$, defined in terms of a uniform probability density $\rho_u$, in the sense that for all possible transitions ${\bf x}\to {\bf x'}$, we have the equality:
\begin{eqnarray}
\label{theoremuniversal}
P^{\rm{univ}}({\bf x}\to {\bf x'})=P({\bf x}\to {\bf x'}|\rho_u).
\end{eqnarray}
}

\noindent From the above theorem, and the previous representation theorem, we can then deduce the following corollary:\\

\vspace{-0.4cm}
\noindent {\bf Corollary}. \emph{If the structure of the set of states of an arbitrary entity is Hilbertian, then the universal measurements performed on that entity are quantum measurements, in the sense that the universal measurements will produce the same values for the outcome's probabilities as those predicted by the Born rule. 
}
\\
\vspace{-0.4cm}

It is important to note that the above doesn't mean that quantum measurements, performed on microscopic entities, would necessarily be universal measurements. This for the time being remains an open question. What we know however is that the huge average involved in a universal measurement is certainly compatible with this interpretation. 

Concerning measurements on cognitive entities, the situation is different, as in this case we have good arguments to affirm that the measurements are the result of an average, considering that they are  performed using a number of different subjects, i.e., of different minds, and that even a single mind, in two different moments, can use in principle different ways of choosing the possible outcomes. Therefore, considering that the Born rule can be interpreted as a universal average, and considering the great success obtained so far by quantum physics in the modeling of human cognition, there are reasons to believe that the measurements performed by human subjects on cognitive entities are in fact universal measurements. This doesn't mean however that they would strictly be quantum measurements, as the structure of the set of states may not be exactly Hilbertian. 

However, what we can say, thanks to our result, is that the model which is behind cognitive measurements does certainly admit a first order approximation, and that whenever the states of the conceptual entity (or the decision process) under investigation is conveniently described by a Hilbert space model, then such a first order approximation precisely corresponds to the quantum mechanical Born rule. And this certainly adds a fascinating piece of explanation as to why quantum mechanics is so successful, also in the description of layers of our reality which are different from the microphysical one.

Before moving to the next session, where we will emphasize the fundamental difference between the GTR-model and the UTR-model, when sequential measurements are considered, we would like to conclude the present one by addressing a possible objection, which is the following. Gleason's theorem, as usually understood, tells us that the Born rule is the only possible measure to be used when computing probabilities in a Hilbert space. Therefore, when the state space is Hilbertian, the Born rule seems to be the only possible measure, and one may wonder about the relevance of the Corollary enunciated in the previous section, stating that, within a Hilbert space, universal measurements deliver the same probabilities as those predicted by the Born rule. 

To answer this objection, it is important to understand that Gleason's theorem doesn't tell us that, if we consider a different measure than the Born rule, we cannot remain anymore inside a Hilbert state space. We certainly cannot do it if we want that this different measure, similarly to the Born rule, only depends on the initial and final states. However, if we relax such condition, then Gleason's theorem doesn't apply anymore, and we can consider all sorts of ``parameters-dependent'' probability measures within a given Hilbertian state space. Now, when considering all these different probability measures, associated with all the possible $\rho$ in our GTR-model, we can certainly also consider their (universal) average, and when we do so what we obtain is that such average gives back a uniform $\rho_u$, in turn associated with the Born rule. 

This provides a possible explanation of why Gleason's theorem imposes us the Born rule as the only consistent rule to be used within a Hilbert space: it would be so because, by asking that the probability measure only depends on the initial state and the final state (or final subspace), it indirectly points to the existence of a physical mechanism that would make such probability measure independent of any other parameter, and our point is that such mechanism is precisely the fact that, in practice, measurements are averages over different measurements, so that the Born rule becomes an effective probability measure obtained by averaging over the different $\rho$-dependent probabilities measures, with the effect that the obtained average measure will only depend on the initial and final states. 

There is another point to emphasize: Gleason's theorem only exists for Hilbert spaces. However, it is not impossible to imagine that generalized Gleason's theorems could also be proved for more general state spaces, pointing to generalized Born rules. These generalized Born rules would be associated, according to our theorem, to uniform $\rho_u$. In quantum mechanics we know that the structure of the state space is Hilbertian, and therefore the uniform $\rho_u$ are precisely those associated with the Born rule. In cognitive systems we haven't yet identified the structure of the set of states, but if our hypothesis regarding the fact that cognitive measurements are (universal) averages over different measurements is correct, we know that uniform $\rho_u$ should be used, and consequently, if the state space is close to a Hilbert space, such uniform $\rho_u$ will be close to the Born rule.

\vspace{-0.4cm}
\section{Non-Kolmogorovian \& non-Hilbertian structures}
\label{nonhilbert}
\vspace{-0.4cm}

In the previous sections  we have defined and analyzed a very general notion of measurement, that we have called \emph{universal measurement}, describing a situation of lack of knowledge which does not refer only to the pure measurement interactions which are  actualized in an unpredictable way, but more generally to the different ``ways of choosing'' these pure measurements, as defined by the different a priori possible probability densities $\rho$. In this section, we want to explicitly show that universal measurements, with their averages, do genuinely describe a deeper level of potentiality, with respect for instance to uniform measurements, in the sense that  when we consider measurements characterized by different $\rho$, we are truly considering a vaster class of measurements, associated with different probability models.

It is indeed natural to ask if a measurement characterized by a non-uniform probability density $\rho$, would really be  fundamentally different than a measurement described by a uniform probability density $\rho_u$. Now, from a physical point of view, it is perfectly clear that, given a state ${\bf x}$, if we perform a series of measurements with a uniform hypermembrane, and then we repeat them with a non-uniform one, different values for the probabilities of the outcomes will be obtained. However, one could  object that these two measurements are in fact structurally equivalent, considering that the UTR-model is already a ``universal probabilistic machine'' \citep{AertsSozzo2012a}. This because, given a non-uniform $\rho$, and a particle in the state ${\bf x}$, it is always possible to find a new $\rho$-dependent state ${\bf x}^{\rho}$, such that $P({\bf x}\to \hat{\bf{x}}_{i}|\rho)=P({\bf x}^{\rho}\to \hat{\bf{x}}_{i}|\rho_u)$, for all $i\in I_N$. In other terms, in the UTR-model it is always possible to fully incorporate into the state of the particle the effects of a non-uniform $\rho$, and therefore, from that perspective, a uniform probability density and a non-uniform one appear to describe the same  measurement situation.

But this  equivalence between the UTR-model, defined in terms of a uniform probability distribution $\rho_u$, and what we have called the GTR-model, which uses general probability distributions $\rho$, is such only if we limit the discussion to a single measurement situation. Indeed, it is when more than a single measurement is considered, and probabilities associated with sequential or conditional measurements are calculated, that the greater structural richness of the GTR-model can be fully revealed. It is precisely the purpose of the present section to show, by means of suitable counter examples, that when probabilities are defined by means of non-uniform $\rho$, and multiple measurements are considered, the obtained probabilities do not necessarily fit into a Kolmogorovian or Hilbertian model, so that the GTR-model does truly generalize the UTR-model and the Hilbert-model. 

To point out the fundamental difference between uniform probability densities (compatible with the quantum mechanical Born rule) and non-uniform ones, we first need to introduce some additional structure. Indeed, it is important to observe that the UTR and GTR models, defined on a single simplex, do not possess per se enough structure to describe in a complete way the different possible states of an entity, in a general experimental setting. Indeed, the UTR and GTR models provide a description only of what we may call a ``naked measurement,'' i.e., of that ``potentiality region'' of contact between the ``outer'' states of the entity under investigation and the ``inner'' states of the measuring apparatus. But a general measurement context contains more information than just that associated with these ``naked'' hidden-measurement interactions, like for instance, in physics, that describing the orientation of the measuring instruments in the three-dimensional Euclidean space (e.g., the orientation of the magnetic field gradient in a typical Stern-Gerlach spin experiment). So, to obtain a more complete and articulated description of the measurement process, this additional information must also be included in the mathematical description of the states of the entity. 

What we are here emphasizing is that in physics probabilities are defined by the combination of two important aspects: the first one is the probability model itself, which originates from the way the different measurement interactions are each time selected, when outcomes are collected; this aspect is described in the GTR-model by the given of a probability distribution $\rho$, which defines the way an elastic hypermembrane breaks; the second aspect is the one coming from the specific spatial orientation and/or location of the measuring apparatuses (polarizers, Stern-Gerlach magnets, etc.) and is usually also incorporated in the way states are defined. 

As far as we know, this second aspect has not  been yet clearly identified and formalized in the description of measurement situations in cognitive experiments. However, it is certainly a relevant aspect  in that ambit as well. To better explain what we mean, let us consider the typical modeling of a concept in cognition (a similar reasoning can also be made for the modeling of a `decision situation'), such as the concept \emph{Fruit}. Even before considering any specific measurement, we can affirm that the concept \emph{Fruit} can exist in different states. There is of course the \emph{ground state} of the concept, corresponding to the idealized situation where the concept presents itself in its ``neutral'' form, i.e., not under the influence of any specific context; but countless excited states can also be described. These excited states can be prepared by simply exposing the concept to the influence of a specific context \citep{AertsGabora2005a, GaboraAerts2002}. For instance, when the concept \emph{Fruit} is combined in the phrase \emph{This is a very juicy fruit}, its state is very different than when in its ground state. 

This difference can be observed when the concept is submitted to a measurement, during which a human mind is asked to select a good exemplar of it, among a given number of possible choices (corresponding to the different possible outcomes of that specific measurement). The difference between the ground state \emph{Fruit} and the excited state \emph{This is a very juicy fruit},  then manifests in the fact that certain examplars -- like \emph{Orange}, \emph{Pineapple} and \emph{Grapes} -- will be chosen much more frequently when the concept is in its excited ``juicy'' state, rather than in its ground state. 

To be a little more specific, let us  denote by $\Sigma$ the state-space of a conceptual entity -- in our example the concept \emph{Fruit} -- i.e., the set of all its possible states, which is clearly a very big set, considering that it contains all possible concept combinations involving that concept. Different possible measurements $e^{\rho}_{\{1\}\cdots\{N\}}$ (for simplicity, we limit our discussion to non-degenerate ones) can obviously be conceived and performed on it,  differing not only in terms of the  possible choices of a $\rho$, but also with regard to the number of outcomes, and the specific states characterizing these outcomes. Now, given a state $s\in\Sigma$, a series of measurements $e^{\rho}_{\{1\}\cdots\{N\}}$ on that state will produce $N$ different probabilities $p_1,\dots,p_N$, and according to (\ref{rhoprobabilitymeasurement}) these probabilities are representable, in the GTR-model, by considering a suitable point ${\bf x}_s$ in a $(N-1)$-dimensional simplex $S_{N-1}$, which is the simplex associated with that specific measurement. This means that we can infer the existence of a map ${\cal M}$ which, to each possible state $s$ of the entity under consideration, and to each possible measurement $e^{\rho}_{\{1\}\cdots\{N\}}$, associates a $(N-1)$-dimensional point ${\bf x}_s$ on a simplex $S_{N-1}$, in such a way that the probabilities of that measurement can be deduced: 
\begin{eqnarray}
&&{\cal M}:\Sigma \rightarrow S_{N-1}\nonumber\\
&&\phantom{{\cal M}:} \, \,\, s \mapsto {\cal M}(s)={\bf x}_s.
\end{eqnarray}

So, according to the above, we can in principle distinguish two different stages in a measurement. The first stage corresponds to that deterministic process during which the state of the concept is first ``brought into contact'' with the `decision context,' which then produces the indeterministic choice of a specific exemplar. In the physical realization of the GTR-model that we have proposed in this article, and in \citet{AertsSassolideBianchi2014a}, this corresponds to  the movement through which the particle reaches the sticky hypermembrane. This first stage of the measurement is represented by the action of the above mentioned map  ${\cal M}$, i.e., it  corresponds to the positioning of the particle in the right location on $S_{N-1}$. When such location is reached, the second stage of the measurement, corresponding to the collapse of the hypermembrane, can then happen, thus producing one of the possible outcomes. 

What we are here evidencing is that contexts can work both deterministically and indeterministically \citep{AertsGabora2005a, AertsGabora2005b}, and that when we describe a measurement in very general terms, both aspects will be present. The deterministic aspect corresponds to that stage of a measurement during which the subject first puts itself in the suitable `decision context,'  creating in this way a sort of `cognitive tension' (represented by the elasticity of the membrane). 
Once this stage is completed, i.e., once the state of the concept is brought into the right position, `ready for a choice to be made,' the second stage of the measurement (that above we have called the ``naked'' measurement) can be implemented, so producing a specific outcome-state, resulting from the sudden and unpredictable reduction of such cognitive tension. 

Now, as we said, and as far as we know, the general description of this ``double stage'' process has not yet been incorporated in more articulated quantum models of cognition and decision, at least not for an arbitrary number $N$ of outcomes, and it is certainly our goal to analyze this problem in future research. But what we wanted to emphasize here is the importance of the interplay between these two different stages, in the characterization of the probability model which describes the different possible measurements that one can perform on a given conceptual (or microscopic) entity. And the purpose of the present section, as already mentioned, is to show this explicitly, in the simple situation where the measurements considered can only have two possible outcomes. 

In physics, this is precisely the situation of typical spin-$1/2$ measurements, where different relative orientations of the Stern-Gerlach apparatus, with respect to an incoming electron prepared in a given spin state, are considered, defining in this way different measurements. In this ambit, the state-space $\Sigma$ is a sphere (the so-called Poincar\'e or Bloch sphere), and the different two-outcome measurements can be represented by 1-simplexes which can have different orientations inside of it. As we will see, the map ${\cal M}$ will then correspond to the ``movement of fall'' of the point particle, initially on the surface of the sphere, onto the 1-simplex elastic band of the measurement under consideration; and when considering this additional state-space structure, it is easy to show that non-uniform probability densities $\rho$ cannot anymore be considered equivalent, from a probabilistic viewpoint, to uniform probability density $\rho_u$.

\begin{figure}[!ht]
\centering
\includegraphics[scale =.8]{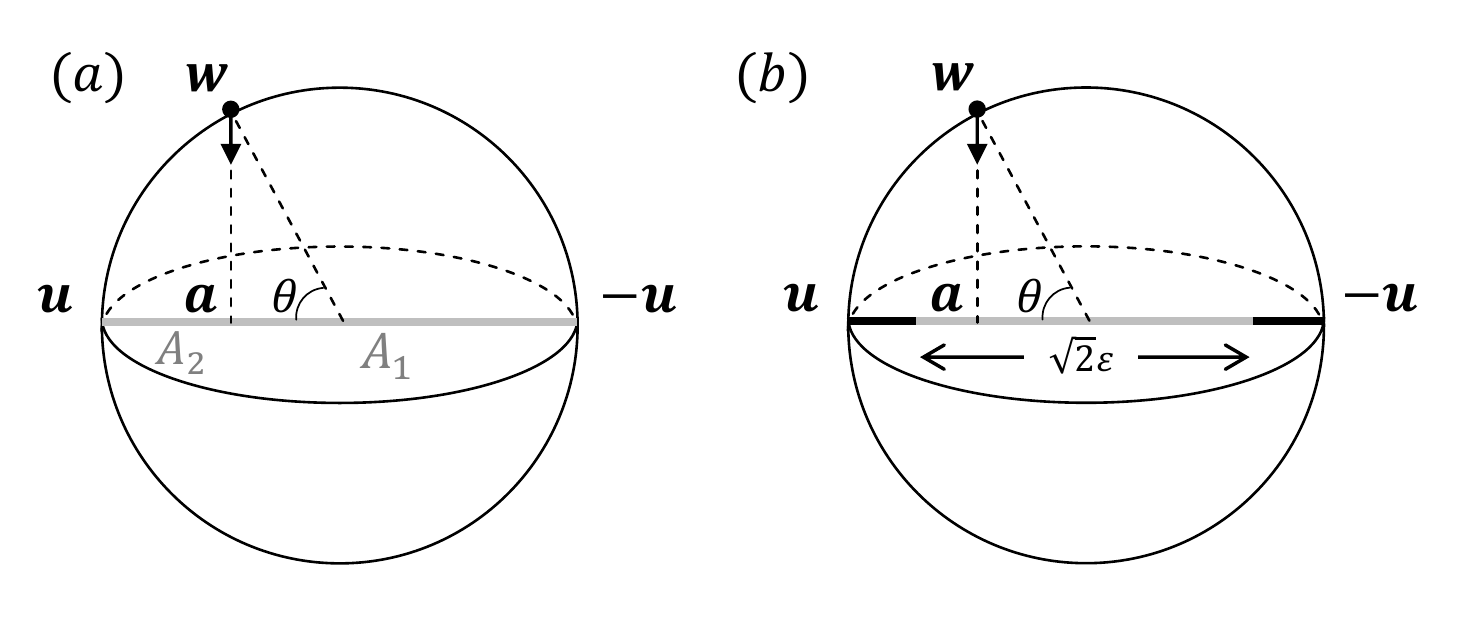}
\caption{$(a)$: The sphere-model, consisting of an empty sphere $S^2$ on which a point particle is located (here at point {\bf w}). A measurement $e^\rho({\bf u})$ is performed by means of a $\rho$-elastic band stretched along the two end-points ${\bf u}$ and $-{\bf u}$. Once the particle has fallen orthogonally onto the elastic (so defining the two regions $A_1$ and $A_2$), it sticks onto it, at point ${\bf a}$, and when the latter breaks, two outcomes are possible: the particle is drawn to point ${\bf u}$ (outcome $1$), or to point $-{\bf u}$ (outcome $2$). In Figure $(b)$, a measurement $e^{\rho_\epsilon}({\bf u})$ is represented, with a $\rho_\epsilon$-elastic which can uniformly break inside a central region of length $\sqrt{2}\epsilon$ (the grey segment), with $\epsilon\in [0,1]$, and is perfectly unbreakable outside of it (the two black lateral segments).}
\label{SphereModel}
\end{figure}

So, we consider now the physical situation of a one-dimensional elastic band (the $N=2$, two-outcome case) embedded in the structure of a $2$-sphere $S^2$, of radius ${1\over \sqrt{2}}$, centered at the origin of the Cartesian coordinate system in ${\mathbb R}^{3}$; a situation known as the \emph{sphere-model}\footnote{Note that in the usual description of the sphere-model, the sphere is of unit radius. Here we choose a sphere of radius ${1\over \sqrt{2}}$, to establish a clearer correspondence with the GTR-model.} \citep{Aertsetal1997b}, which in turn is a generalization of the so-called $\epsilon$\emph{-model} \citep{Aerts1998, Aerts1999b, SassolideBianchi2013a, SassolideBianchi2013b}. More precisely, the initial state of the material point particle is now a point\footnote{We have changed notation, using the letter ${\bf w}$ instead of the letter ${\bf x}$, to generically denote states, to make clear that these now belong to the $2$-sphere $S^2$, and not to the $1$-simplex $S_1$. Also, note that the origin of the coordinate system now coincides with the center of the sphere.} ${\bf w}\in S^2$. A two-outcome measurement $e^\rho({\bf u})\equiv e^{\rho}_{\{1\}\{2\}}({\bf u})$ consists then in the following operations (see Fig.~\ref{SphereModel} $(a)$): a $\rho$-elastic band is stretched and fixed at two diametrically opposite points ${\bf u}$ and $-{\bf u}$ of the $2$-sphere, and once this is done, the particle in state ${\bf w}$ ``falls'' orthogonally onto the elastic and sticks on it, at point ${\bf a}$. Then, the elastic breaks, and draws the particle either to point ${\bf u}$ (outcome $1$), or to point $-{\bf u}$ (outcome $2$). 

Introducing the angle $\theta\equiv \theta ({\bf w},{\bf u})$ between the two vectors ${\bf w}$ and ${\bf u}$, we have that ${\bf a} = \cos\theta\, {\bf u}$, with $\cos\theta = 2\, {\bf w}\cdot {\bf u}$, so that the two regions $A_2$ and $A_1$, generated by the presence of the particle on the elastic, correspond to the two intervals $[{\cos\theta\over\sqrt{2}},{1\over\sqrt{2}}]$ and $[-{1\over\sqrt{2}},{\cos\theta\over\sqrt{2}}]$, respectively. Therefore, the probabilities to obtain outcomes $1$ and $2$, respectively, when $e^\rho({\bf u})$ is performed, are given by the two integrals: 
\begin{eqnarray}
P({\bf w}\to {\bf u}|\rho) = P(A_1|\rho) = \int_{-{1\over\sqrt{2}}}^{{\cos\theta\over\sqrt{2}}}\rho(z)dz,\\
P({\bf w}\to -{\bf u}|\rho) = P(A_2|\rho) = \int_{{\cos\theta\over\sqrt{2}}}^{{1\over\sqrt{2}}}\rho(z)dz.
\label{sphereprob}
\end{eqnarray}

Let us consider a special choice of a non-uniform probability density, which is the following: $\rho_\epsilon (z)\equiv {1\over \sqrt{2}\epsilon}\chi_{[-{\epsilon\over\sqrt{2}},{\epsilon\over\sqrt{2}}]}(z)$, where $\chi_{[-{\epsilon\over\sqrt{2}},{\epsilon\over\sqrt{2}}]}$ is the characteristic function of the interval $[-{\epsilon\over\sqrt{2}},{\epsilon\over\sqrt{2}}]$, and $\epsilon$ is a positive parameter which can take values between $0$ and $1$ (see Fig.~\ref{SphereModel} $(b)$).
Then, if we perform a measurement $e^{\rho_\epsilon}({\bf u})$ associated with a $\rho_\epsilon$-elastic oriented along direction ${\bf u}$, we observe that if the initial state $\bf{w}$ of the particle is such that, when it falls orthogonally onto the elastic, it lands on its left (resp. right) unbreakable segment, then the probability for outcome $1$ is equal to $1$ (resp. $0$), and the probability for outcome $2$ is equal to $0$ (resp. $1$). On the other hand, if the initial state $\bf{w}$ of the particle is such that it lands on its central uniformly breakable segment (this is the situation depicted in Fig.~\ref{SphereModel} $(b)$), then the probabilities for the two outcomes can be calculated by simply considering the ratio between the length of the piece of breakable elastic between the particle and the end-point, divided by the total length $\sqrt{2}\epsilon$ of the breakable segment.

Putting all this together, one obtains the following formula: 
\begin{eqnarray}
P({\bf w}\to \pm{\bf u}|\rho_\epsilon) =\delta_{\pm,-1}\Theta(-\cos\theta -\epsilon) + \delta_{\pm,+1}\Theta(\cos\theta -\epsilon)+\frac{1}{2}\left(1\pm {\cos\theta\over \epsilon}\right)\chi_{[-\epsilon,\epsilon]}(\cos\theta),
\label{probability2b}
\end{eqnarray}
\noindent where $\Theta$ denotes the Heaviside step function (equal to 1 when the argument is positive and equal to 0 otherwise),  and $\delta$ denotes the Kronecker delta (equal to 1 when the two indices are the same and equal to 0 otherwise). Clearly, in the limit $\epsilon\to 0$, $\rho_\epsilon(z) \to \rho_0(z) = \delta(z)$, i.e., the elastic becomes only breakable in its middle point and therefore corresponds to a pure measurement, with no fluctuations, and the third term of (\ref{probability2b}) vanishes, so that one recovers an almost classical situation (almost because for $\cos\theta = \pm\epsilon$ we are in a situation of unstable equilibrium). 

\begin{figure}[!ht]
\centering
\includegraphics[scale =.8]{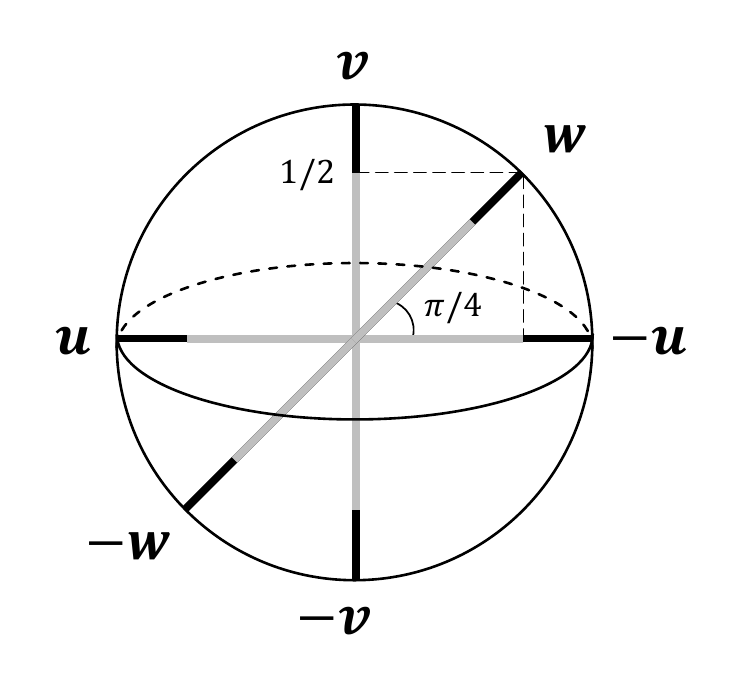}
\caption{Three different measurements $e^{\rho_\epsilon}({\bf u})$, $e^{\rho_\epsilon}({\bf v})$ and $e^{\rho_\epsilon}({\bf w})$, represented by three $\rho_\epsilon$-elastic bands oriented along directions ${\bf u}$, ${\bf v}$ and ${\bf w}$, respectively. The angle between ${\bf w}$ and ${\bf v}$ is ${\pi\over 4}$, and the angle between ${\bf v}$ and ${\bf u}$ is ${\pi\over 2}$. The unbreakable segments of the elastics are in black color, the uniformly breakable segments in grey color, and the picture corresponds to the choice $\epsilon = {\sqrt{2}\over 2}$ (the breakable segments are of length $1$). 
\label{threeMes}}
\end{figure}

In the opposite limit $\epsilon \to 1$, i.e., $\rho_\epsilon(z)\to \rho_1(z) ={1\over \sqrt{2}}\chi_{[-{1\over\sqrt{2}},{1\over\sqrt{2}}]}(z)\equiv \rho_u(z)$, the first two terms of (\ref{probability2b}) vanish, and the third one tends to the quantum probabilities describing a Stern-Gerlach measurement with a spin-$1/2$ microscopic entity: 
\begin{eqnarray}
\label{probability3}
P({\bf w}\to \pm{\bf u}|\rho_u) = \frac{1}{2}\left(1\pm\cos\theta\right) =\delta_{\pm,-1}\sin^2{\frac{\theta}{2}} + \delta_{\pm,+1}\cos^2{\frac{\theta}{2}}.
\end{eqnarray}
On the other hand, for $\epsilon\neq 0,1$, we are in a situation of intermediate knowledge, where depending on the value taken by $\theta$, we can either predict with certainty the outcome (if $|\cos\theta|>\epsilon$), or not (if $|\cos\theta|<\epsilon$). This situation cannot be fitted, in general, neither into a classical probability model nor into a quantum one, which is what we are now going to show. For this, we assume that $\epsilon \in [0, {\sqrt{2}\over 2}]$, and we consider three different measurements: $e^{\rho_\epsilon}({\bf u})$, $e^{\rho_\epsilon}({\bf v})$ and $e^{\rho_\epsilon}({\bf w})$, with an angle of ${\pi\over 4}$ between ${\bf w}$ and ${\bf v}$, and an angle ${\pi\over 2}$ between ${\bf v}$ and ${\bf u}$ (see Fig.~\ref{threeMes}).
\\
\vspace{-0.3cm}
\\
\emph{The probability model is non-classical}
\\
We start by showing that the joint probabilities of sequential measurements cannot be fitted into a classical probability model, by means of a reasoning of \emph{reductio ad absurdum}. For this, we observe that if $U$, $V$ and $W$ are three arbitrary events (elements of a $\sigma$-algebra of subsets of a sample space), and $U^c$, $V^c$ and $W^c$ are their complement, then, according to classical probability theory (obeying Kolmogorovian axioms), if $P$ is a probability measure, then the following three equalities hold: 
\begin{eqnarray}
\label{setEqualities}
&&P(V\cap W) = P(U\cap V\cap W)+P(U^c \cap V\cap W)\label{eq1},\\
&&P(U\cap W) = P(U\cap V\cap W)+P(U \cap V^c \cap W)\label{eq2},\\
&&P(U^c\cap V) = P(U^c \cap V\cap W)+P(U^c \cap V \cap W^c).\label{eq3}
\end{eqnarray}
Eq. (\ref{eq1}) minus (\ref{eq2}) gives:
\begin{eqnarray}
\label{setEqualities2}
P(U^c \cap V\cap W) = [P(V\cap W) - P(U\cap W)] + P(U \cap V^c \cap W),
\end{eqnarray}
and considering that the last term in (\ref{setEqualities2}) is positive, we have:
\begin{eqnarray}
\label{setEqualities3}
P(U^c \cap V\cap W)\geq [P(V\cap W) - P(U\cap W)]. 
\end{eqnarray}
Also, from (\ref{eq3}), we can deduce that 
\begin{eqnarray}
\label{setEqualities4}
P(U^c \cap V\cap W)\leq P(U^c\cap V), 
\end{eqnarray}
and putting together (\ref{setEqualities3}) and (\ref{setEqualities4}), we obtain:
\begin{eqnarray}
\label{setEqualities5}
[P(V\cap W) - P(U\cap W)]\leq P(U^c \cap V\cap W)\leq P(U^c\cap V), 
\end{eqnarray}
which implies that: 
\begin{eqnarray}
\label{setEqualities6}
[P(V\cap W) - P(U\cap W)]\leq P(U^c\cap V). 
\end{eqnarray}
So, if we can prove that equality (\ref{setEqualities6}) is violated by the $\epsilon$-model (here with $\epsilon \in [0,{\sqrt{2}\over 2}]$), then we have also proved that it cannot be fitted into a classical probabilistic model. For this, we assume that the particle is in state ${\bf w}$, i.e., an eigenstate of measurement $e^{\rho_\epsilon}({\bf w})$. This means that if we perform $e^{\rho_\epsilon}({\bf w})$, we will obtain outcome $1$ with certainty, i.e., $P(\to{\bf w}|{\bf w}) =1$. The probability of obtaining outcome {\bf v}, when we perform $e^{\rho_\epsilon}({\bf v})$, with the particle in state ${\bf w}$, is also equal to $1$, i.e., $P(\to {\bf v}|{\bf w}) =1$, since the particle falls on the unbreakable segment with end point {\bf v}, as one can easily check on Fig.~\ref{threeMes}. Therefore, the joint probability that, given the particle in state ${\bf w}$, we obtain first state ${\bf w}$, then state ${\bf v}$, in a sequential measurement, is $P(\to{\bf w}\to {\bf v}|{\bf w})=P(\to{\bf w}|{\bf w})P(\to{\bf v}|{\bf w})=1\cdot 1 = 1$. On the other hand, since $P(\to{\bf u}|{\bf w}) =0$ (see Fig.~\ref{threeMes}), we also have that $P(\to{\bf w}\to {\bf u}|{\bf w})=P(\to{\bf w}|{\bf w})P(\to{\bf u}|{\bf w})=1\cdot 0 = 0$. Also, considering that when we perform $e^{\rho_\epsilon}({\bf u})$, a particle in state ${\bf v}$ falls exactly on the middle point of the elastic, we have $P(\to{\bf v}\to -{\bf u}|{\bf w})=P(\to{\bf v}|{\bf w})P(\to -{\bf u}|{\bf v})=1\cdot {1\over 2} = {1\over 2}$. Thus, being that $[1-0]>{1\over 2}$, we obtain:
\begin{eqnarray}
\label{setEqualities7}
[P(\to{\bf w}\to {\bf v}|{\bf w}) -P(\to{\bf w}\to {\bf u}|{\bf w})]>P(\to{\bf v}\to -{\bf u}|{\bf w}),
\end{eqnarray}
which is clearly a violation of (\ref{setEqualities6}).

Note that, for simplicity, we have here constructed a counter example only using probabilities of sequential measurements, interpreted as joint probabilities. An alternative \emph{ex absurdum} proof, using instead conditional probabilities, can also be constructed, considering a situation where there is an additional lack of knowledge about the  state of the particle, described by a uniform probability distribution. Then, choosing again three suitable measurements, one can show that Bayes' rule for conditional probabilities is violated. The calculation of conditional probabilities, with an additional mixture of states, is however much more involved, and for the proof we refer the interested reader to \citet{Aerts1986} and \citet{Aerts1995}.
\\
\vspace{-0.3cm}
\\
\emph{The probability model is non-quantum}
\\
Let us now show that the probabilities of sequential measurements cannot be fitted into a quantum probability model. We denote by $|\psi_{\bf u}\rangle$ and $|\psi_{-{\bf u}}\rangle$ the two orthonormal eigenstates associated with the two outcomes of measurement $e^{\rho_\epsilon}({\bf u})$, respectively. Similarly, we denote by $|\psi_{\bf v}\rangle$ and $|\psi_{-{\bf v}}\rangle$ those associated with measurement $e^{\rho_\epsilon}({\bf v})$, and by $|\psi_{\bf w}\rangle$ and $|\psi_{-{\bf w}}\rangle$ those associated with $e^{\rho_\epsilon}({\bf w})$. According to the standard quantum formalism, if the initial state of the system is $|\psi_{\bf w}\rangle$, we can write:
\begin{eqnarray}
\label{setEqualities8}
&&1= P(\to{\bf w}\to {\bf v}|{\bf w}) = |\langle\psi_{\bf v}|\psi_{\bf w}\rangle|^{2} |\langle\psi_{\bf w}|\psi_{\bf w}\rangle|^{2} = |\langle\psi_{\bf v}|\psi_{\bf w}\rangle|^{2},\\
&&1= P(\to{\bf w}\to -{\bf u}|{\bf w}) = |\langle\psi_{-\bf u}|\psi_{\bf w}\rangle|^{2} |\langle\psi_{\bf w}|\psi_{\bf w}\rangle|^{2} = |\langle\psi_{-\bf u}|\psi_{\bf w}\rangle|^{2}\label{setEqualities9},\\
&&{1\over 2}= P(\to{\bf v}\to {\bf u}|{\bf w}) =|\langle\psi_{\bf u}|\psi_{\bf v}\rangle|^{2} |\langle\psi_{\bf v}|\psi_{\bf w}\rangle|^{2} = |\langle\psi_{\bf u}|\psi_{\bf v}\rangle|^{2},
\label{setEqualities10}
\end{eqnarray}
where for (\ref{setEqualities9}) we have used $P(\to{\bf w}\to -{\bf u}|{\bf w}) = P(\to{\bf w}| {\bf w})P(\to -{\bf u}|{\bf w}) = 1\cdot 1=1$, and for (\ref{setEqualities10}) we have used (\ref{setEqualities8}) and $P(\to{\bf v}\to {\bf u}|{\bf w})= P(\to{\bf v}| {\bf w})P(\to {\bf u}|{\bf v}) = 1\cdot {1\over 2}={1\over 2}$. Clearly, (\ref{setEqualities8}) implies that $\langle\psi_{-\bf w}|\psi_{\bf v}\rangle=0$, (\ref{setEqualities9}) that $\langle\psi_{\bf u}|\psi_{\bf w}\rangle=0$, and (\ref{setEqualities10}) that $\langle\psi_{\bf u}|\psi_{\bf v}\rangle \neq 0$. Therefore:
\begin{eqnarray}
\label{setEqualities11}
0\neq \langle\psi_{\bf u}|\psi_{\bf v}\rangle =\langle\psi_{\bf u}|\psi_{\bf w}\rangle\langle\psi_{\bf w}|\psi_{\bf v}\rangle+\langle\psi_{\bf u}|\psi_{-\bf w}\rangle\langle\psi_{-\bf w}|\psi_{\bf v}\rangle=0,
\end{eqnarray}
which is clearly a contradiction, showing that there does not exist a two-dimensional Hilbert space model such that the above sequential transition probabilities can be described in this Hilbert space. 

It is not difficult to analyze the above three measurements also for values $\epsilon\in [{\sqrt{2}\over 2}, 1]$, and show that (\ref{setEqualities7}) continues to hold, which means that for all values $\epsilon\in [0,1]$, the $\epsilon$-model is non-Kolmogorovian. Using Accardi-Fedullo inequalities \citep{AccardiFedullo1982}, it is also possible to show that in general no Hilbert space model exists, except for the uniform $\epsilon =1$ case \citep{Aertsetal1999a}.

It follows from the above analysis that when non-uniform probability densities and different measurements are considered (i.e., different simplexes), the overall probability model will be in general non-Hilbertian and non-Kolmogorovian, with the Hilbertian structure corresponding only to the special case of a uniform probability density $\rho_u$. Of course, each single measurement can be described within a single Kolmogorovian probability space (in accordance with the fact that the UTR-model is a universal single-measurement probabilistic machine\footnote{When only a single measurement is considered, also the Hilbert-model is of course Kolmogorovian.}), but when different measurements are considered, the overall probability model will in general be non-Kolmogorovian and non-Hilbertian. Indeed, the actual structure of the model depends on the specific choice of the $\rho$ describing how the elastic structures break (and on the structure of the set of states which are  mapped to points  on the simplexes), and this means that the GTR-model constitutes a truly more general theoretical framework, able to extend its ``structural reach'' far beyond that of the UTR-model and Hilbert-model.

But the above sphere-model example also shows that, although non-uniform $\rho$ can give rise to non-Hilbertian probability models, as soon as universal measurements are considered, i.e., averages over all possible probability densities $\rho$ are taken, being that these averages will give back an effective condition of uniform fluctuation, the Hilbertian structure for the set of states can also possibly be recovered.

A last remark is in order. Although in our model all the probabilities are calculated within a Lebesgue measure context, it would not be correct to think of it in classical terms (for instance as a Markov chain), when used to describe sequential measurements. Let us note in this regard that in a Markov chain, and more generally in a classical stochastic process, all the random variables are defined with respect to a same probability space, and in particular to a same $\sigma$-algebra of events. On the other hand, the transition probabilities arising from the application in sequence of the UTR-model (or more generally of the GTR-model), cannot be fit into a single algebra of events, even though the individual probabilities are calculated, in each single measurement, by means of a classical Lebesgue measure. The reason for this is that in our model each measurement brings its own specific input of randomness, in a way that makes the different measurements mutually incompatible, which is not the case for a classical stochastic process, where all the randomness comes from a lack of knowledge about the state of the considered entity.
\\
\vspace{-0.3cm}
\\
\emph{The general $N$-outcome situation}
\\
The previous analysis is only valid for the two-outcome (qubit) situation ($N=2$). Therefore, one might ask if the Corollary we have enunciated in the previous section, expressing the equivalence between universal measurements and quantum measurements (as described by the Born rule), when the structure of the set of states is Hilbertian, also applies in general, or is just a two-dimensional anomaly. Indeed, a full ``membrane modelization'' of the quantum measurement process was only provided in the two-outcome situation, by means of the sphere-model, while for a general $N$-outcome situation we have only provided a description of the ``naked'' part of the measurement process. The following question therefore arises: Is it possible to provide a complete description of a quantum measurement (also valid for degenerate observables), in the case of a general finite-dimensional Hilbert space? 

Undoubtedly, only if such a description can be obtained we can affirm that the UTR probability model, when the state space is Hilbertian, becomes perfectly equivalent to the probability model generated (within the same state space) by the Born rule, and this not only for single measurements, but also for arbitrary sequential measurements, in accordance with the projection postulate. To put the same question in different terms: Is it possible to generalize the sphere-model for a quantum entity having an arbitrary number $N$ of dimensions? Also, can we inscribe and orient, within a suitable geometric structure, the different $(N-1)$-simplexes associated with all possible observables, and specify in this ambit the nature of the deterministic map ${\cal M}$ that, to each state in ${\cal H}_N={\mathbb C}^N$, associates a specific point on the $(N-1)$-simplex describing the measurement under consideration, in a way that systematically works for any possible choice of measurement (that is, for any possible choice of physical observable)? 

When this article, and its companion \citep{AertsSassolideBianchi2014a}, were submitted for publication, this question was still an open one. However, very recently we succeeded in fully determining the map ${\cal M}$, and the  mathematical structure that naturally contains all the possible measurement simplexes associated with a $N$-dimensional quantum entity, thus obtaining a general hidden-measurements solution to the  quantum measurement problem, for general $N$-outcome experiments \citep{AertsSassolideBianchi2014b}.

Let us just briefly mention here that for more than two outcomes, the three-dimensional Bloch-sphere representation is replaced by a $(N^2-1)$-dimensional generalized Bloch-sphere representation, but only a convex portion of such generalized sphere can be filled with states. However, similarly to the two-dimensional situation, the map ${\cal M}$ still describes a point particle ``orthogonally falling'' onto the measurement simplex. When the measurement is degenerate, the process is a little more articulated, in the sense that there is not only a deterministic process through which the point particle enters into contact with the membrane, but also a possible final deterministic movement, following the membrane's collapse, bringing the point particle back to a maximum distance from the center of the simplex, in accordance with L\"uders-von Neumann projection formula.
 
This hidden-measurements description of a general quantum measurement is quite involved mathematically speaking, as it uses the properties of the generators of the group $SU(N)$, the special unitary group of degree $N$, and the representation of quantum states as density operators. Therefore, we refer the interested reader to \citep{AertsSassolideBianchi2014b}, for the mathematical details. As regards the present analysis, what is important to keep in mind is that a more general description exists, so that the equivalence between the UTR-model and the Born rule, as expressed in the Corollary of Sec.~\ref{Non-uniform}, has a very general validity.

\vspace{-0.4cm}
\section{Conclusion}
\label{Conclusion}
\vspace{-0.4cm}

To conclude, we briefly summarize the results we have presented in this article. We have introduced an abstract model -- the UTR-model --  which like the Hilbert-model of quantum mechanics is able to describe all possible probabilities in a given single measurement context.  We have also shown that the UTR-model admits a simple physical realization, in which measurements are described as actions performed on a point particle by means of special, uniformly breakable elastic hypermembranes. These hypermembranes can also be understood as the exemplification of psychological processes during which a human subject reduces the tension of a mental state of equilibrium, between different competing alternatives, by means of a random perturbation, breaking in this way the symmetry of the equilibrium and bringing the system into a more stable and less tensional condition.   

The UTR-model illustrates in a detailed way what possibly happens during a measurement, when the initial state in which the entity is prepared collapses into a final state, showing that the process can be understood in terms of an actualization of hidden potential interactions, which (almost) deterministically bring the entity into its final state. We have also shown that a symmetrical interpretation is also possible, where the potentiality is the result of a uniform mixture of state, instead of a uniform mixture of pure measurements. 

Despite the already great generality of the UTR-model, we have motivated the introduction of an even more general model, that we have called the GTR-model, which employs non uniform hypermembranes (non-uniform probability densities), describing all possible kinds of  quantum and quantum-like measurements, including the pure deterministic and pure solipsistic ones. 

In the much ampler structural ambit of the GTR-model, we have then considered the possibility of performing uniform averages over all possible choices of non-uniform probability densities $\rho$, i.e., averages over all possible probability models which can in principle be actualized in a given experimental context. This uniform average is what we have called a \emph{universal measurement}, to which we have given a mathematically precise and physically transparent definition, as the limit of an average over finite structures, so bypassing (i.e., solving) the well known difficulties of so-called Bertrand paradox
\citep{AertsSassolideBianchi2014c}   

A universal measurement corresponds to the situation were not only there is lack of knowledge about which specific hidden interaction is actualized during the measurement, but also about the way this interaction is selected. This means that a universal measurement corresponds to a situation of maximum possible lack of knowledge, involving a double level of randomization: one related to the choice of the measurement itself, and one related to the actualization of the hidden deterministic interaction within that chosen measurement. 

The result we have outlined -- which we will prove in \citep{AertsSassolideBianchi2014a} -- is that the ``huge'' randomization of a universal measurement produces exactly the same  numerical values for the transition probabilities as those delivered by a uniform probability density, i.e., by the Lebesgue measure on the simplexes; and since the latter is compatible with the Born rule, this also means that quantum measurements can actually be understood as universal measurements, thus adding an important piece of explanation regarding the effectiveness of the quantum model in so many different ambits, such as that of human cognition and decision.

These results are significant for cognitive science, considering that the hypothesis of the conceptual-mental layer of individual subjects participating in experiments to contain deep variations is a plausible one. Hence, our analysis indicates that it may be  possible  that more careful and numerous cognitive experiments would reveal that the average which is at play is not as huge and systematic as the quantum mechanical average, so that the  statistics of outcomes may actually deviate from that predicted by an effective uniform $\rho_u$, and therefore also from the Born rule. This possible deviation, however, can only be evidenced when considering experimental situations involving more than a single measurement (sequential measurements), as we have shown in Sec.~\ref{nonhilbert}.

More specifically, in the special case of two-outcome measurements, and in the ambit of the so-called sphere-model (see \citep{AertsSassolideBianchi2014b} for its recent $N$-dimensional generalization), which takes into consideration the full geometry of the Hilbert state space, we have shown that the GTR-model is a much more general framework than the UTR-model, in the sense that it can generate probabilities that cannot fit into a Kolmogorovian or Hilbertian model. This means that universal measurements, when considered from the viewpoint of sequential measurements, are not only averages over different probability values, but also averages over different probability models.   

In any case, even if the effective uniform $\rho_u$, i.e., the ``Lebesgue rule'' on the simplexes (which becomes exactly the Born rule when the structure of the set of states is Hilbertian) will prove in the  future not to always be the good rule to apply to infer the statistics of outcomes of  cognitive experiments, or other experiments in other regions of reality, it follows from the present analysis that, in the absence of a  specific knowledge about how an experiment is specifically conducted, it certainly constitutes the best possible prediction, as it corresponds to a `first order non-classical theory,' expressing a condition of maximum lack of knowledge and control.

Let us conclude by saying that the results presented in this article, which will be further explored in its second part \citep{AertsSassolideBianchi2014a}, are of special interest also for physics. Our results contain a potential deep explanation for what happens during a quantum measurement in micro-physics, and further more refined experiments might explore this explanation. Indeed, if this explanation is correct, the fact that  quantum mechanics describes a statistics of outcomes that goes along with a uniform at-random choice between any arbitrary type of manipulation that changes an initial state of the entity under study into a final state, in a way that we don't know anything about the mechanism of this change of state, reveals then, that the micro-layer of our physical reality is characterizable by a much deeper level of potentiality than was initially expected. In the sense that, apparently, all possible measurements are also equipossible measurements, which are in principle actualizable and actualized in the laboratory. If we have not realized this so far, it could be because these measurements remain hidden, and their different individual statistics of outcomes are fused together, in a unique statistics, equivalent to that delivered by the Born rule.  
\\
\vspace{-0.3cm}
\\
{\bf Acknowledgements} We are grateful to the anonymous referees for their numerous insightful comments, which have contributed in considerably improving the presentation of our work. We are also grateful to Jerome Busemeyer, for his careful reading of the manuscript and for highlighting the psychological mechanism which is inherent in our membrane models, as well as for having suggested a particularly apt name for them.

\appendix

\vspace{-0.4cm}
\section{The uniform probability density}
\label{Uniform}
\vspace{-0.4cm}

To calculate the probability (\ref{transitionprobabilitiesnondeg}) we follow here the derivation in \citet{Aerts1986}. We observe that the Lebesgue measure of $S_{N-1}$ is $\mu_L(S_{N-1})= {\sqrt{N}\over (N-1)!}$  (for $N=2$, it corresponds to the length $\sqrt{2}$ of the line segment $S_1$, for $N=3$, to the area ${\sqrt{3}\over 2}$ of the equilateral triangle $S_2$, for $N=4$, to the volume ${1\over 3}$ of the regular tetrahedron $S_3$, and so on). Therefore,  (\ref{transitionprobabilitiesnondeg}) becomes:
 \begin{eqnarray}
\label{rhoprobabilitymeasurement2}
P({\bf x}\to \hat{\bf{x}}_{i})  = {(N-1)!\over \sqrt{N}}\mu_L(A_i).
\end{eqnarray}
To calculate $\mu_L(A_i)$, we can use the generalization, for a convex hull, of the well-known formula for the computation of the area of a triangle, as the product of the length of its base times its height, times ${1\over 2}$, which in the case of the $(N-1)$-dimensional convex hull $A_i$ becomes: 
\begin{eqnarray}
\label{Ai}
\mu_L(A_i) &=& {1\over N-1}\mu_L(S^i_{N-2})h^i({\bf x}) =  {1\over N-1} {\sqrt{N-1}\over (N-2)!}h^i({\bf x})\nonumber\\
&=& {\sqrt{N-1}\over (N-1)!}h^i({\bf x}),
\end{eqnarray}
where $S^i_{N-2}$ is the $(N-2)$-dimensional simplex generated by the $N-1$ orthonormal vectors $\hat{\bf x}_1, \dots, \hat{\bf x}_{i-1},\hat{\bf x}_{i+1}, \dots\hat{\bf x}_{N+1}$, and $h^i({\bf x})$ is the smallest Euclidean distance between ${\bf x}$ and $S^i_{N-2}$. To calculate $h^i({\bf x})$, we observe that any point of $S^i_{N-2}$ can be written as ${\bf y}^{i} =\sum_{j=1\atop j\neq i}^{N}y_j^i \hat{\bf x}_j$, with $\sum_{j=1\atop j\neq i}^{N+1}y_j^i =1$, so that the vector ${\bf x}-{\bf y}^{i}$, on the line connecting ${\bf x}$ and ${\bf y}^{i}$, is given by:
\begin{eqnarray}
\label{vector}
{\bf x}-{\bf y}^{i} = \sum_{j=1\atop j\neq i}^{N}(x_j - y_j^i)\hat{\bf x}_j + x_i\hat{\bf x}_i.
\end{eqnarray}
To find the ${\bf y}^{i}$ for which the distance $\|{\bf x}-{\bf y}^{i}\|$ is minimal, i.e., for which $\|{\bf x}-{\bf y}^{i}\|=h^i({\bf x})$, we observe that for such vector ${\bf x}-{\bf y}^{i}$ must be orthogonal to all vectors of the form $\hat{\bf x}_j-\hat{\bf x}_k$, with $j,k\neq i$, that is, $({\bf x}-{\bf y}^{i}|\hat{\bf x}_j-\hat{\bf x}_k)=0$, for all $j,k\neq i$. This  implies that $x_j-y_j^i =
x_k-y_k^i$, for all $j,k\neq i$, so that all the differences $x_j-y_j^i $, $j\neq i$, must be equal to a same constant $c$. To determine $c$, we use $\sum_{j=1\atop j\neq i}^{N}y_j^i =1$ and $\sum_{j=1\atop j\neq i}^{N}x_j = 1-x_i$, to write $\sum_{j=1\atop j\neq i}^{N}(x_j - y_j^i) = -x_i$. Therefore, $(N-1)c =-x_i$, i.e., $c=-{x_i\over N-1}$. Eq.~(\ref{vector}) then becomes: 
\begin{eqnarray}
\label{vector-bis}
{\bf x}-{\bf y}^{i} = -{x_i\over N}\sum_{j=1\atop j\neq i}^{N}\hat{\bf x}_j + x_i\hat{\bf x}_i,
\end{eqnarray}
so that 
\begin{eqnarray}
\label{vector-tris}
h^i({\bf x})=\|{\bf x}-{\bf y}^{i}\| = \sqrt{\left({x_i^2\over N-1}\right)^2 + x_i^2}=\sqrt{N \over N-1}\, x_i, 
\end{eqnarray}
and inserting (\ref{vector-tris}) into (\ref{Ai}), gives $\mu_L(A_i) = {\sqrt{N}\over (N-1)!} x_i$, so that (\ref{rhoprobabilitymeasurement2}) yields:
\begin{eqnarray}
\label{probability-itris}
P({\bf x}\to \hat{\bf x}_{i})=x_i, \quad i\in I_N.
\end{eqnarray}

\vspace{-0.4cm}
\section{Mixed states: the 3-outcome case}
\label{Mixed}
\vspace{-0.4cm}

\begin{figure}[!ht]
\centering
\includegraphics[scale =.6]{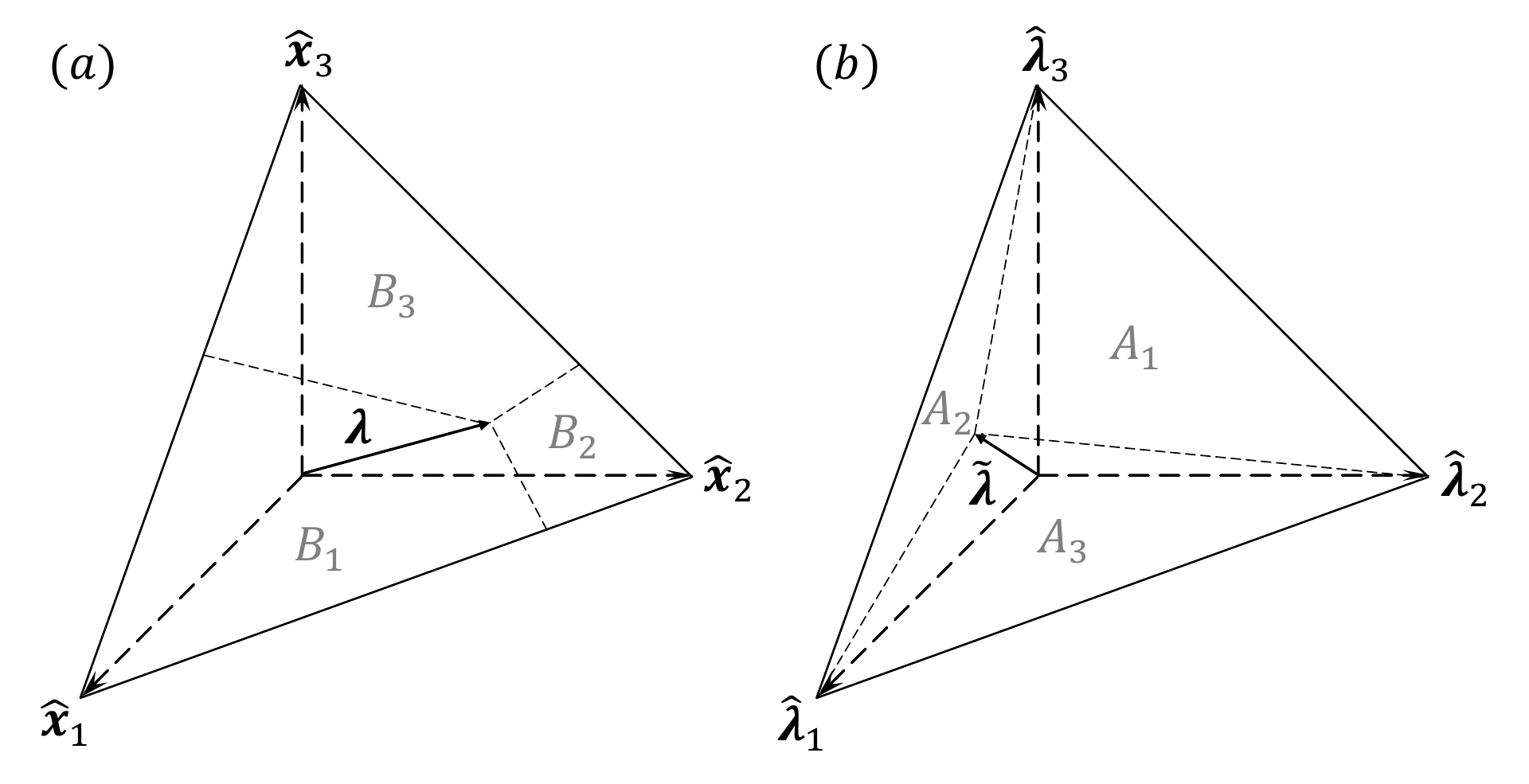}
\caption{$(a)$: A $2$-dimensional triangular membrane which can only break  at a single point \mbox{\boldmath$\lambda$} (contrary to Fig.~\ref{triangolo}, the membrane is now represented in white, to mark the fact that, with the exception of a single point, it is unbreakable), with the three quadrilateral regions $B_i$ consisting of those initial states that are all drawn to the same outcome $\hat{\bf x}_i$, $i=1,2,3$. $(b)$: The same pure measurement represented as a point \mbox{\boldmath${\tilde \lambda}$} in the 2-dimensional simplex generated by the final states of the measuring entity, with the region $A_i$ corresponding to those initial states of the point particle that change the state of the measuring entity into the final states $\hat{\mbox{\boldmath$\lambda$}}_i$, $i=1,2,3$.
\label{triangolobis}}
\end{figure}

We consider the situation of a pure measurement in the presence of a uniform mixture of states, in the case $N=3$ (see Sec.~\ref{RepresentingProbabilities}). As it can be checked on Fig.~\ref{triangolobis} $(a)$, to obtain the states of the point particle that are drawn to a same  outcome $\hat{\bf{x}}_i$, $i=1,2,3$, one has to prolong the lines that connect the vertex points of $S_2$ to  \mbox{\boldmath$\lambda$}, to the opposite sides of the 2-simplex. In this way, one obtains three disjoint quadrilateral regions $B_i$, $i=1,2,3$, and by calculating their areas\footnote{To determine the area of the quadrilateral regions $B_i$, one has to observe that they are the sum of two triangles, sharing two vertices. The coordinates of these vertices being known, their respective area $\Delta$ can be easily calculated, using the formula:
\begin{equation}
\Delta={1\over 2}\sqrt{\left|
\begin{array}{ccc}
y_1 & z_1 & 1 \\
y_2& z_2 & 1 \\
y_3 & z_3 & 1 \end{array}
\right|^2 +\left|
\begin{array}{ccc}
z_1 & x_1 & 1 \\
z_2& x_2 & 1 \\
z_3 & x_3 & 1 \end{array}
\right|^2+\left|
\begin{array}{ccc}
x_1 & y_1 & 1 \\
x_2& y_2 & 1 \\
x_3 & y_3 & 1 \end{array}
\right|^2},
\end{equation}
where the triples $(x_i,y_i,z_i)$, $i=1,2,3$, are the coordinates of the three vertices of the triangle.} as a function of the components of \mbox{\boldmath$\lambda$}, one finds the probabilities $P(\to\hat{\bf{x}}_1)={\mu_L(B_1)\over \mu_L(S_1)}={\lambda_2\lambda_3(1+\lambda_1)\over (1-\lambda_2)(1-\lambda_3)}$, $P(\to\hat{\bf{x}}_2)={\mu_L(B_2)\over \mu_L(S_1)}={\lambda_1\lambda_3(1+\lambda_2)\over (1-\lambda_3)(1-\lambda_1)}$, and $P(\to\hat{\bf{x}}_3)={\mu_L(B_3)\over \mu_L(S_1)}={\lambda_1\lambda_2(1+\lambda_3)\over (1-\lambda_1)(1-\lambda_2)}$.
As for the two-outcome case considered in Sec.~\ref{RepresentingProbabilities}, we can represent these probabilities by considering an additional 2-simplex, with the state of the apparatus now described by a vector \mbox{\boldmath${\tilde \lambda}$} generating three triangular regions $A_i$ (see Fig.~\ref{triangolobis} $(b)$), so that the probabilities can again be written in the simpler form $P(\mbox{\boldmath${\tilde \lambda}$} \to\hat{\mbox{\boldmath$\lambda$}}_i)={\mu_L(A_i)\over \mu_L(S_2)}=\tilde{\lambda}_i$, $i=1,2,3$, and the connection with the quantum mechanical Born rule is realized by describing the state of the measuring entity by means of the Hilbert space vector $|\phi\rangle = \sqrt{\tilde{\lambda}_1}e^{i\beta_1}|b_1\rangle +\sqrt{\tilde{\lambda}_2}e^{i\beta_2}|b_2\rangle +\sqrt{\tilde{\lambda}_3}e^{i\beta_3}|b_3\rangle\in {\mathcal H}_3$.

\end{document}